\documentclass[a4paper,11pt]{article}
\pdfoutput=1 

\usepackage{jcappub}
\usepackage[toc,page]{appendix}
\usepackage{subcaption}
\usepackage{enumerate}
\usepackage{amsmath}

\usepackage{graphicx}                
\usepackage{multirow}
\usepackage{braket}
\usepackage[T1]{fontenc} 
\usepackage[table]{xcolor}
\usepackage{float}

\definecolor{verde}{rgb}{0,0.5,0}

\def\be{\begin{equation}}
\def\ee{\end{equation}}
\def\bea{\begin{eqnarray}}
\def\eea{\end{eqnarray}}
\def\be{\begin{equation}}
\def\ee{\end{equation}}
\def\ba{\begin{eqnarray}}
\def\ea{\end{eqnarray}}

\hypersetup{pdftitle={\ref{FIX ME}}
}
\usepackage{bm}

\title{\boldmath Primordial correlators from multi-point propagators}

\author[a]{Andrea Costantini,}
\author[a,b]{Laura Iacconi}
\author[a]{and David J. Mulryne}
\affiliation{$^{a}$Astronomy Unit, Queen Mary University of London, \\
Mile End Road, London, E1 4NS, UK}
\affiliation{$^{b}$Institute of Cosmology \& Gravitation, University of Portsmouth, \\
Burnaby Road, Portsmouth, PO1 3FX, UK}

\emailAdd{a.costantini@qmul.ac.uk}
\emailAdd{l.iacconi@qmul.ac.uk}
\emailAdd{d.mulryne@qmul.ac.uk}

\abstract{A key step in the comparison between inflationary predictions and cosmological observations is  the computation of primordial correlators. 
Numerical methods have been developed that overcome some of the difficulties arising in analytical calculations when the models considered are complex.
The \texttt{PyTransport} package, which implements the transport formalism, allows computation of the tree-level 2- and 3-point correlation functions for multi-field models with arbitrary potentials and a curved field space. 
In this work we investigate an alternative numerical implementation of the transport approach, based on the use of transfer ``matrices'' called \textit{multi-point propagators} (MPP).
We test the novel MPP method, and extensively compare it with the traditional implementation of the transport approach provided in \texttt{PyTransport}. 
We highlight advantages of the former, discussing its performance in terms of accuracy, precision and running time, as well as dependence on the number of e-folds of sub-horizon evolution and tolerance settings. 
For topical ultra-slow-roll models of inflation we show that MPPs (i) precisely track the decay of correlators even when \texttt{PyTransport} produces erroneous results, (ii) extend the computation of squeezed bispectra for squeezing values at least one decade beyond those attainable with \texttt{PyTransport}.
}

\begin{document}
	\maketitle
	\flushbottom
\section{Introduction}
\label{sec: intro}

Cosmological inflation is a phase of accelerated expansion in the very early universe. 
Initially proposed to address the main problems of the standard Hot Big Bang cosmology~\cite{Starobinsky:1980te, PhysRevD.23.347, PhysRevLett.48.1220, Hawking:1981fz, Linde:1981mu, Linde:1983gd}, it also provides a mechanism for explaining the origin of the large-scale structure we observe in the Universe today~\cite{Liddle_Lyth_2000}.
For these reasons, inflation has become the leading paradigm to describe the early universe and, as such, a key phase of the standard model of cosmology.

Cosmological inflation can be realised through the dynamics of a single scalar field, minimally coupled to gravity.
On a sufficiently flat region of its potential, the field slowly rolls, leading to a period of accelerated expansion.
All models which realise this mechanism go under the name of single-field slow-roll (SFSR) inflation, and the scalar field is usually called the inflaton. 
Quantum fluctuations produced during SFSR inflation lead to nearly scale-invariant and almost Gaussian primordial curvature perturbations.  
This prediction is in 
agreement with Cosmic Microwave Background (CMB) measurements~\cite{Planck:2018jri}, which constrain the statistics of primordial curvature perturbations on large scales. 

While primordial non-Gaussianities (PNG) produced within SFSR inflation are slow-roll suppressed~\cite{Gangui:1993tt, Maldacena:2002vr, Celoria:2018euj}, 
alternative models of inflation (such as single-field models with non-standard kinetic terms or higher-derivative interactions, or multi-field models) can predict higher values of non-Gaussianity, still compatible with the latest constraints~\cite{Planck:2019kim}. 
Detecting, or further constraining PNG, therefore remains one of the most important objectives of upcoming CMB and large-scale structure (LSS) surveys, one that will shed light on the physical processes in action during inflation (see e.g. the Review~\cite{Achucarro:2022qrl}). 

CMB and LSS experiments constrain perturbations on large scales ($10^{-3}\lesssim k/\text{Mpc}^{-1}\lesssim 10^{-1}$). 
Nevertheless, the statistics of primordial perturbations on much shorter scales is also of great current interest, and much less constrained~\cite{Gow:2020bzo, LISACosmologyWorkingGroup:2025vdz}.
On these scales, deviations from the predictions of simple SFSR models are therefore possible.
For example, the variance of primordial curvature perturbations could be large enough on small scales to lead to significant production of primordial black holes (PBH)~\cite{Carr:1974nx} -- a possible explanation for a fraction of dark matter~\cite{Green:2024bam}. 
In this context, non-Gaussianity of the curvature perturbation plays an important role~\cite{Ferrante:2022mui, Gow:2022jfb}.

Key quantities which carry information on physical processes at play during inflation are the correlation functions of the primordial curvature\footnote{While in this work we focus on the description of curvature perturbations produced during inflation, also primordial gravitational waves are a generic prediction of inflationary theories, see e.g. the Review~\cite{Guzzetti:2016mkm}.} perturbation, the \textit{primordial correlators}. 
These allow to us compare observations and theoretical predictions, and therefore to discriminate among different inflationary scenarios. 
For example, Gaussian perturbations are fully characterised by the 2-point correlation function, whose Fourier transform is the power spectrum. 
Deviations from Gaussianity encode additional information on the primordial universe physics. 
The leading deviation from the Gaussian distribution is given by the 3-point correlation function, whose Fourier transform is the bispectrum. 

In this work we return to the issue of methods to compute primordial correlators. 
We work in a very general setting, considering models with an arbitrary number of fields\footnote{This approach is motivated by the need to accommodate inflation within high-energy theories, e.g. string theory, which predicts the existence of more than one degree of freedom~\cite{Baumann:2014nda}.}, arbitrary analytical forms for the fields' potential, and possibly a non-trivial field-space metric. 
Primordial correlators are routinely computed by applying methods developed in the context of Quantum Field Theory on curved spacetime~\cite{Birrell:1982ix}. 
For example, the In-In Schwinger Keldysh formalism~\cite{Weinberg:2005vy, Weinberg:2006ac, Maldacena:2002vr} has led to  some remarkable results, such as the computation of non-Gaussianity for single-~\cite{Maldacena:2002vr, Seery:2005wm} and multi-field inflation~\cite{Seery:2005gb}.
However, these (analytical) calculations are often complicated to perform, and the complexity grows with the number of fields and the perturbative order one needs to attain. 
These issues can be (at least partially) overcome by switching to numerical methods.
In early-universe cosmology, numerical packages to compute the 2- and 3-point functions at tree-level have been developed, including: \texttt{BINGO}~\cite{Hazra:2012yn}, which employs an integral formalism to compute the bispectrum for single-field models; numerical implementations of the \textit{transport approach}~\cite{Mulryne:2009kh, Mulryne:2010rp,  Seery:2012vj,Anderson:2012em, Mulryne:2013uka, Dias:2016rjq},
such as \texttt{mTransport}~\cite{Dias:2015rca}, \texttt{CppTransport}~\cite{Seery:2016lko} and \texttt{PyTransport}~\cite{Mulryne:2016mzv, Ronayne:2017qzn}, where the latter two allow computation of the bispectrum for multi-field models with curved field space (see also the complimentary code \texttt{CosmoFlow}~\cite{Werth:2023pfl, Pinol:2023oux, Werth:2024aui}, which applies the transport formalism to the effective field theory framework).
The essence of the transport approach is to set up coupled ordinary differential equations for the correlation functions of the inflationary model phase-space variables. 
Once these have been evolved up to, e.g., the end of inflation, they can converted to the primordial correlators of the curvature perturbation by means of a gauge transformation. 
The transport approach is expected to be completely equivalent to the In-In formalism, and this has been shown explicitly at tree-level~\cite{Mulryne:2013uka,Dias:2016rjq}. 
A key feature of the transport approach to the computation of primordial correlators is its convenience for numerical implementation, as numerical solutions of differential equations are often more stable than, e.g., numerical integration. 

In this work our aim is to explore an alternative implementation of the transport approach, which 
relies on transfer ``matrices'' called \textit{multi-point propagators} (MPP).
MPP measure the (2-point, 3-point, \textit{etc.}) correlation between non-linearly evolved fields and their value(s) at some initial time.
They have been previously considered in different contexts, 
for example Bernardeau \textit{et al} introduced them in Ref.~\cite{Bernardeau:2008fa} to study gravitational instability in cosmological perturbation theory.
Within the transport approach, the importance of MPPs resides in the fact that they provide a formal --albeit implicit-- solution to the transport equations for the cosmological correlators, as first discussed in Ref.~\cite{Seery:2012vj}. 
In practice, MPPs obey their own equations of motion, and therefore constitute an alternative to the direct evolution of the phase-space correlators performed in the traditional implementation of the transport approach~\cite{Dias:2015rca, Seery:2016lko, Mulryne:2016mzv, Ronayne:2017qzn}.

In Sec.~\ref{sec: transport approach primer} we review the transport approach to primordial correlators, in its traditional implementation.
We introduce multi-field inflation models with curved field-space in Sec.~\ref{sec: background}, and discuss the definition of covariant scalar perturbations in Sec.~\ref{sec: scalar perturbations}. 
We present the transport system in Sec.~\ref{sec: hamilton to transport}, and show how its solutions can be used to build primordial correlators in Sec.~\ref{sec: define correlators}.
In Sec.~\ref{sec: mpp} we introduce MPPs and their transport equations, spelling out the difference between the traditional implementation of the transport approach and the novel one with MPPs. 
In Sec.~\ref{sec: MPP at work} we test our approach against the traditional \texttt{PyTransport} package, and study its performance in terms of accuracy, precision and running time, as well as dependence on the number of e-folds of sub-horizon evolution and tolerance settings. 
We test our new code using a variety of inflationary models: double quadratic (Sec.~\ref{sec: double quadratic}), axion-quartic (Sec.~\ref{sec: axion-quartic}), single-field model with a potential feature (Sec.~\ref{sec: single field with feature}), two-field model with non-trivial field-space metric (Sec.~\ref{sec: two field non-canonical}). 
We then select two ultra-slow-roll models~\cite{Kinney:2005vj, Dimopoulos:2017ged, Pattison:2018bct}, topical in light of their small-scale phenomenology, and show that MPPs (i) precisely track the decay of correlators when \texttt{PyTransport} fails (Sec.~\ref{sec: USR and large-scale bispectrum}), (ii) extend the computation of squeezed bispectra for squeezing values at least one decade beyond those attainable with \texttt{PyTransport} (Sec.~\ref{sec: USR and squeezed bispectrum}). 
In Sec.~\ref{sec: discussion} we draw our conclusions, and discuss the flexibility of application of the MMP approach, for example to construct a wider range of observables including loop corrections to primordial correlators. 

A pre-release of newly developed MPP code can be found \href{https://github.com/Zer095/PyTransport3.0}{here}, where it is included within an update to the \texttt{PyTransport} package.
We will follow up with an accompanying paper~\cite{Costantini_in_prep} providing a detailed guide to the new tools.  

\section{A transport approach primer}
\label{sec: transport approach primer}
We review here the traditional implementation of the transport approach to primordial correlators, see e.g. Ref.~\cite{Ronayne:2017qzn} for the case of multiple field inflation with non-trivial field-space geometry. 
The model and background evolution are introduced in Sec.~\ref{sec: background}, and in Sec.~\ref{sec: scalar perturbations} we discuss covariant field perturbations. 
We summarise the derivation of transport equations for 2- and 3-point phase-space correlators in Sec.~\ref{sec: hamilton to transport}, and show how these are then related to the (tree-level) power spectrum and bispectrum of the primordial curvature perturbation in Sec.~\ref{sec: define correlators}. 

\subsection{Background evolution}
\label{sec: background}
We consider multi-field models of inflation with $\mathcal{N}$ scalar fields, minimally coupled to gravity and with generic field-space metric, $G_{IJ}$, and potential, $V$. 
The action reads 
\begin{equation}
    \mathcal{S} = \frac{1}{2}\int \mathrm{d}^4x\sqrt{-g}\left[ M_{Pl}^2R - G_{IJ}g^{\mu\nu}\partial_\mu\phi^{I}\partial_\nu\phi^{J} - 2V\right]
    \label{eq: multi-field action} \;, 
\end{equation}
where $R$ is the Ricci scalar associated with the spacetime metric $g_{\mu\nu}$, and Roman upper-case indices run over the number of fields. 
From now on we work in units where $M_{Pl} = 1$.

On flat FLRW spacetime, the background evolution is dictated by 
\begin{subequations}
    \label{eq: background evolution}
    \begin{align}
    3 H^2 - \frac{1}{2} G_{IJ}\dot \phi^I \dot \phi^I -V&=0\;, \\
    D_tQ^I + 3H \dot \phi^I +G^{IJ}V_{,J} &=0 \;, 
    \end{align}
\end{subequations}
where an overdot indicates a derivative with respect to cosmic time, and the covariant derivative is defined as $D_tQ^I \equiv \dot Q^I + \Gamma_{J K}^I \dot{\phi}^K Q^J$, with $\Gamma^I_{JK}$ being the connection compatible with the metric $G_{IJ}$. 

\subsection{Scalar perturbations in the spatially-flat gauge}
\label{sec: scalar perturbations}

While tensor perturbations are also produced from inflation, in this work we consider only scalar perturbations. 
In the spatially-flat gauge, the relevant scalar perturbations are in the matter sector. 
These require careful handling when expanding the action~\eqref{eq: multi-field action} to cubic order or higher, as the field perturbations $\delta \phi^I(t,\mathbf{x}) \equiv \phi^I(t,\mathbf{x}) - \phi^I_0(t)$ do not transform covariantly under a change of coordinate in field space.  
Here, $\phi^I_0(t)$ represents the background solution obtained from Eqs.~\eqref{eq: background evolution}. 
The coordinate displacement $\delta \phi^I(t,\mathbf{x})$ and the covariant field perturbation $Q^I(t,\mathbf{x})$ can be related as~\cite{Gong:2011uw}
\begin{equation}
     \delta\phi^{I} = Q^{I} - \frac{1}{2!}\Gamma^{I}_{JK}Q^{J}Q^{K} + \mathcal{O}(Q^3) \;. 
     \label{eq: field to covariant}
\end{equation}
By differentiating Eq.~\eqref{eq: field to covariant}  with respect to time,  one can relate the velocity $\delta \dot \phi^I$ to $Q^I$ and $D_tQ^I$. 

To obtain the quadratic and cubic actions for the covariant field perturbations one then proceeds as follow: (i) perturb the matter sector as described above, and the metric sector in the ADM decomposition~\cite{Arnowitt:1962hi, Misner:1973prb}; (ii) solve for the metric shift and lapse functions by using the energy and momentum constraint equations; (iii) select the spatially-flat gauge, where the metric curvature perturbation, $\zeta(t,\mathbf{x})$, is set to zero. 
Complete expressions for the quadratic and cubic actions can be found in Ref.~\cite{Ronayne:2017qzn}. 

\subsection{From Hamilton equations to transport equations}
\label{sec: hamilton to transport}

The covariant field perturbation $Q^I$ behaves quantum-mechanically in the sub-horizon regime. 
To quantize it, we introduce its canonical conjugate momentum, 
\begin{equation}
     \hat P_I \equiv \frac{\delta \mathcal{S}}{\delta[ D_t \hat Q^I ]} \;, 
     \label{eq: conjugate momentum definition}
\end{equation}
and impose the commutator relation
\begin{equation}
    \label{eq: commutator}
    \left[\hat Q^I(t, \mathbf{k}), \,  \hat P_J(t', \mathbf{k'})\right] = i \,(2\pi)^3 \delta^I_J \, \delta^3(\mathbf{k+k'})  \, \delta (t-t')\;. 
\end{equation}
In the following we drop the hat symbol, with the understanding that $Q^I$ and $P_J$ are quantum fields. 
 
The evolution of the Heisenberg picture fields $Q^I$ and $P_J$ can be obtained from Hamilton's equations\footnote{It is useful to rescale $ P_J$ such that $P_J \longrightarrow a^3 P_J$, where we make use of the same symbol for ease of notation. Eq.~\eqref{eq: hamilton for P} is different from its canonical form due to this rescaling.},
\begin{subequations}
\label{eq: heisenberg fields eom}
\begin{align}
    D_tQ^I &= -i\left[Q^I,\mathcal{H}\right]\;, \\
    \label{eq: hamilton for P}
    D_tP^I &= -i\left[P^I,\mathcal{H}\right] - 3HP^I \;. 
\end{align}
\end{subequations}
Here, the Hamiltonian function of the system, $\mathcal{H}$, is obtained by performing the Legendre transform of the Lagrangian density. 
Up to third order, this yields
\begin{multline}    
\label{eq: hamiltonian}
    \mathcal{H} = \frac{1}{2} \int \mathrm{d}t\; a^3 \Bigg[ \int \mathrm{d}^3k_a \int \mathrm{d}^3k_b\; \mathcal{H}_{ab}(\mathbf{k}_a, \mathbf{k}_b)\, \delta^3(\mathbf{k}_a+\mathbf{k}_b) \,\delta X^a(\mathbf{k}_a)\,\delta X^b(\mathbf{k}_b) \\
    + \int \mathrm{d}^3k_a \int \mathrm{d}^3k_b \int \mathrm{d}^3k_c\; \mathcal{H}_{abc}(\mathbf{k}_a, \mathbf{k}_b,\mathbf{k}_c)\, \delta^3(\mathbf{k}_a+\mathbf{k}_b+\mathbf{k}_c) \,\delta X^a(\mathbf{k}_a)\,\delta X^b(\mathbf{k}_b)\,\delta X^b(\mathbf{k}_c)\Bigg]\;. 
\end{multline}
$\mathcal{H}_{ab}$ is the free part of the Hamiltonian, and the part encoding cubic interactions is $\mathcal{H}_{abc}$. 
Complete expressions for these can be found in Ref.~\cite{Ronayne:2017qzn}.
For compactness, we have introduced the $2\mathcal{N}$-dimensional phase-space vector $\delta X^a = \left(Q^1, \, Q^2,\cdots, P_1, \, P_2,\, \cdots \right)$.

The knowledge of the equations of motion for $Q^I$ and $P_I$ is enough to obtain evolution equations for the expectation values of products of these operators via the Ehrenfest's theorem~\cite{Mulryne:2013uka}. 
In this work we focus on the 2- and 3-point phase-space correlators, which are defined as 
\begin{subequations}
    \begin{align}
    \langle\delta X^a\left(\mathbf{k}_a\right) \delta X^b\left(\mathbf{k}_b\right)\rangle & \equiv (2 \pi)^3 \delta^3\left(\mathbf{k}_a+\mathbf{k}_b\right) \Sigma^{a b}(k_a) \;, \label{eq: phase-space correlator 2pt} \\
    \langle\delta X^a\left(\mathbf{k}_a\right) \delta X^b\left(\mathbf{k}_b\right) \delta X^c\left(\mathbf{k}_c\right)\rangle & \equiv (2 \pi)^3 \delta^3\left(\mathbf{k}_a+\mathbf{k}_b+\mathbf{k}_c\right) B^{a b c}(k_a, k_b, k_c) \;. 
    \label{eq: phase-space correlator 3pt}
    \end{align}
\end{subequations}

They satisfy the \textit{transport equations}
\begin{subequations}
\label{eq: traditional transport equations}
    \begin{align}
    \label{eq: transport Sigma}
    D_t \Sigma^{ab} (k_a) &= {u^a}_c(k_a)\Sigma^{cb}(k_a) + {u^b}_c(k_a)\Sigma^{ac}(k_a) \;,  \\
    \label{eq: transport B}
D_t B^{a b c}\left(k_a, k_b, k_c\right) &=  {u^a}_d\left(k_a\right) B^{d b c}\left(k_a, k_b, k_c\right)+{u^b}_d\left(k_b\right) B^{a d c}\left(k_a, k_b, k_c\right)+{u^c}_d\left(k_c\right) B^{a b d}\left(k_a, k_b, k_c\right) \nonumber\\
& \quad+{u^a}_{d e}\left({k}_a,{k}_b,{k}_c\right) \Sigma^{d b}\left(k_b\right) \Sigma^{e c}\left(k_c\right) \nonumber\\
& \quad +{u^b}_{d e}\left({k}_b,{k}_a,{k}_c\right) \Sigma^{a d}\left(k_a\right) \Sigma^{e c}\left(k_c\right)\nonumber \\
&\quad  +{u^c}_{d e}\left({k}_c,{k}_a,{k}_b\right) \Sigma^{a d}\left(k_a\right) \Sigma^{b e}\left(k_c\right) \;. 
    \end{align}
\end{subequations}
The $u$-tensors appearing in the transport equations can be obtained from the quadratic and cubic Hamiltonian~\cite{Ronayne:2017qzn}. 
The last piece needed to obtain the time-evolution of the correlators, using Eqs.~\eqref{eq: traditional transport equations}, are appropriate initial conditions in the deep sub-horizon regime. 
If the initial time is set sufficiently early, the 2-point correlation function approaches that of a set of massless, uncoupled scalar fields and is therefore model independent. 
For the 3-point function, the In-In formalism is then used to derive analytic expressions in the massless regime, where an analytic calculation becomes tractable, and the expressions derived are valid for all models~\cite{Dias:2016rjq,Ronayne:2017qzn}. 
Note that usually in numerical codes that implement the transport equations, e.g. \texttt{PyTransport}, time is measured in e-folds of expansion, defined as $N\equiv \int\mathrm{d}t \,H$.

\subsection{From phase-space variables to the primoridial curvature perturbation}
\label{sec: define correlators}

Inflation provides the initial conditions for the subsequent cosmological evolution. 
The inflationary predictions of a specific model are encoded into the statistics of $\zeta(t,\mathbf{x})$, the primordial curvature perturbation defined on uniform-density hypersurfaces. 
If $\zeta$ is a Gaussian variable, its statistical properties are fully described by the 2-point correlation function, or equivalently the power spectrum. 
In Fourier space this is defined as 
\begin{equation}
    \langle\zeta(\textbf{k}_1) \zeta(\textbf{k}_2)\rangle \equiv \left(2\pi\right)^3\, \delta^3(\textbf{k}_1 + \textbf{k}_2)\, P_{\zeta}(k_1) \;, 
    \label{eq: zeta power spectrum}
\end{equation}
from which one can introduce the dimensionless scalar power spectrum
\begin{equation}
    \mathcal{P}_{\zeta}(k) \equiv \frac{k^3}{2\pi^2}\, P_{\zeta}(k)\;. 
    \label{eq: zeta dimensionless}
\end{equation}
However, if the inflationary dynamics is not fully described by Gaussian processes, higher-order correlators are needed to describe the statistics of $\zeta$. 
The first deviation from Gaussianity is measured by the 3-point correlation function. 
Its Fourier transform is related to the primordial bispectrum $B_\zeta$, 
\begin{equation}
    \langle\zeta(\textbf{k}_1) \zeta(\textbf{k}_2) \zeta(\textbf{k}_3)\rangle \equiv \left(2\pi\right)^3\, \delta^3(\textbf{k}_1 + \textbf{k}_2 + \textbf{k}_3)\, B_{\zeta}(k_1,k_2,k_3) \;, 
\label{eq: zeta bispectrum}
\end{equation}
from which one defines the reduced bispectrum  
\begin{equation}
        f_\text{NL}(k_1,k_2,k_3) \equiv \frac{5}{6}\frac{B_{\zeta}\left(k_1,k_2,k_3\right)}{P_\zeta(k_1)P_\zeta(k_2)+P_\zeta(k_2)P_\zeta(k_3)+P_\zeta(k_1)P_\zeta(k_3)} \;. 
        \label{eq: fnl}
\end{equation}
The primordial curvature perturbation can be related to the phase-space variables discussed in Sec.~\ref{sec: hamilton to transport} through a gauge transformation~\cite{Dias:2014msa},
\begin{equation}
\label{eq: gauge transformation}
    \zeta(\textbf{k}) = N_a ({k})\delta X^{a}(\textbf{k})
    + \frac{1}{2!} \int \frac{\mathrm{d}^3 k_a}{(2\pi)^3}\,  N_{ab}({k}, {k}_a, |\textbf{k}-\textbf{k}_a|) \, \delta X^{a}(\mathbf{k}_a)\, \delta X^{b}(\textbf{k}-\textbf{k}_a) + \mathcal{O}\left(\delta X^3 \right) \;. 
\end{equation}
For the explicit expressions of $N_{a}$ and $N_{ab}$ see e.g. Ref.~\cite{Dias:2016rjq}. 

The transport system~\eqref{eq: traditional transport equations} dictates the time-evolution of the phase-space 2- and 3-point correlators, $\Sigma^{ab}$ and $B^{abc}$. 
The 2-point phase-space correlator is a complex quantity, and it can be decomposed into its real and imaginary parts, $\Sigma^{ab} = \Sigma^{ab}_\text{Re} + i\Sigma^{ab}_\text{Im}$. 
However, only the real part is required to compute the 2-point correlator of $\zeta$. 
In principle, also the 3-point correlator can be complex, but its imaginary part vanishes at tree level, so it does not contribute to the calculation.
Thanks to Eq.~\eqref{eq: gauge transformation} these can then be used to build $P_\zeta$ and $B_\zeta$,
\begin{subequations}
    \label{eq: gauge transformations for Pz and Bz}
    \begin{align}
    P_\zeta (k) &= N_a(k) \, N_b(k) \, \Sigma_\text{Re}^{ab}(k) \;,  \\
    B_\zeta (k_1,k_2,k_3)  &= N_a(k_1) \, N_b(k_2) \, N_c(k_3) \, B_\text{Re}^{abc}(k_1,k_2,k_3)\nonumber \\
    & \quad \quad + \left(N_a(k_1)\, N_b(k_2)\, N_{c}(k_3,|\mathbf{k_2}+\mathbf{k}_3|, k_2)\, \Sigma^{ac}_\text{Re}(k_1)\, \Sigma_\text{Re}^{bd}(k_2) + \, 2\,  \text{cyclic}  \right) \;. 
\end{align}
\end{subequations}
To summarise, the transport formalism is a differential approach to the computation of primordial correlators. 
The set of steps outlined here and in the previous subsections constitute the backbone of the implementation of the transport approach in state-of-the-art numerical codes, such as \texttt{PyTransport}.
Note that the transport approach is fundamentally different from the In-In (or Schwinger-Keldysh) formalism~\cite{Weinberg:2005vy, Weinberg:2006ac, Maldacena:2002vr}. 
Due to the non-conservation of energy on a cosmological background, the In-In method entails computation of time integrals over the whole duration of inflation. 
The transport approach recasts the problem of time-evolution into differential equations, and is highly amenable to a numerical implementation, whilst yielding the same results when comparison is possible~\cite{Dias:2016rjq,Ronayne:2017qzn}. 

\section{MPP approach to transport equations}
\label{sec: mpp}
In this Section we discuss an alternative implementation of the transport approach.   
This relies on transfer ``matrices'', called \textit{multi-point propagators} (MPPs)~\cite{Seery:2012vj}.
In early universe settings, MPPs measure how late-time phase-space fields depend on early-time ones. 
For the purpose of computing the power spectrum and bispectrum of $\zeta$, it suffices to define the first two MPPs~\cite{Seery:2012vj},
\begin{subequations}
    \label{eq: MPPs defs}
    \begin{align}
    \frac{\partial \delta X^a(N, \textbf{k}_1)}{\partial \delta X^i(N_0, \textbf{k}_2)} & \equiv \delta^3( \textbf{k}_1 - \textbf{k}_2)\, \Gamma^{a \; (N_0, N)}_i(k_1) \;, \label{eq: mpp2 def}  \\
    \frac{\partial \delta X^a(N,\textbf{k}_1)}{\partial \delta X^i(N_0,\textbf{k}_2)\;\partial\delta X^j(N_0,\textbf{k}_3)} & \equiv \delta^3( \textbf{k}_1 - \textbf{k}_2 - \textbf{k}_3)\, \Gamma^{a \; (N_0, N)}_{ij}(k_1, k_2, k_3) \;. \label{eq: mpp3 def}
\end{align}
\end{subequations}
The MPPs $\Gamma^{a \; (N_0, N)}_i(k_1)$ and $\Gamma^{a \; (N_0, N)}_{ij}(k_1, k_2, k_3)$ are momentum-dependent objects, that carry upper (lower) indices to indicate late-time (early-time) phase-space variables, as well as time labels indicating the early and late times, $N_0$ and $N>N_0$ respectively. 
Note that the first two MPPs depend only on the magnitude of the momenta due to the isotropy and homogeneity of the background.
Eqs.~\eqref{eq: MPPs defs} show that MPPs measure the correlation between non-linearly evolved fields at time $N$ and their value(s) at $N_0$~\cite{Bernardeau:2008fa}. 

MPPs can be used to Taylor-expand the late-time phase-space variables in terms of early-time ones. 
Up to second-order we have
\begin{multline}
    \delta X^a \left(N, \textbf{k}\right) = \Gamma^{a \; \left(N_0, N\right) }_i(k) \,  \delta X^i\left(N_0, \textbf{k}\right) + \\
    + \frac{1}{2!} \int \frac{\mathrm{d}^3{k}_1}{\left(2\pi\right)^3} \, \Gamma^{a \; \left(N_0, N\right)}_{i j}(k,k_1,|\mathbf{k}-\mathbf{k}_1|) \,\delta X^i\left(N_0,\mathbf{k}_1\right) \, \delta X^j\left(N_0,\mathbf{k}-\mathbf{k}_1\right) \;. 
    \label{eq: mpp field}
\end{multline}
Note that here we label the phase-space variables also with time, as the emphasis is on their time dependence. 
By using Eqs.~\eqref{eq: heisenberg fields eom} and the definition~\eqref{eq: mpp field}, one obtains an equation of motion for the first two MPPs~\cite{Seery:2012vj, Mulryne:2013uka}
\begin{subequations}
    \label{eq: transport eqs for MPPs}
    \begin{align}
    \label{eq: mpp2 eom}
    D_t \, \Gamma_{i}^{a \;\left(N_0, N\right)}\left(k\right) &= {u^a}_b(k)\, \Gamma_{i}^{b \;\left(N_0, N\right)}(k) \;, \\
    \label{eq: mpp3 eom}
    D_t\, \Gamma_{i j}^{a \; \left(N_0, N\right)}(k_1,k_2,k_3) &= {u^a}_b(k_1)\, \Gamma_{i j}^{b \;\left(N_0, N\right)}(k_1,k_2,k_3) \nonumber\\
    &\quad  + {u^a}_{bc}(k_1,{k}_2,{k}_3)\, \Gamma_{i}^{b \;\left(N_0, N\right)}(k_2)\, \Gamma_{ j}^{c \;\left(N_0, N\right)}(k_3) \;. 
\end{align}
\end{subequations}
Form the definitions~\eqref{eq: MPPs defs} it follows that the initial conditions are
\begin{equation}
    \Gamma^{a \;\left(N_0,N_0\right)}_{i} = \delta_{i}^a \quad \text{and}\quad \Gamma^{a \;\left(N_0,N_0\right)}_{i j} = 0 \;. 
    \label{eq: mpp ics}
\end{equation}
MPPs obtained from solving Eqs.~\eqref{eq: transport eqs for MPPs} can then be used to propagate the 2- and 3-point phase-space correlators from $N_0$ to $N$. 
This is achieved by employing the expansion in Eq.~\eqref{eq: mpp field} in Eqs.~\eqref{eq: zeta power spectrum} and~\eqref{eq: zeta bispectrum}, leading to 
\begin{subequations}
    \label{eq: use MPP for sigma and B at late times}
    \begin{align}
        \label{eq: mpp sigma}
        \Sigma^{ab}\left(N,\, k\right) &= \Gamma_{ i}^{a \;\left(N_0, N\right)}(k)\, \Gamma_{j}^{b \;\left(N_0, N\right)}(k)\, \Sigma^{ij}\left(N_0,\, k\right)\;,  \\
        \label{eq: mpp bispectrum}
        B^{abc}\left(N, k_1,k_2,k_3\right) &= \Gamma_{i}^{a \;\left(N_0, N\right)}\left(k_1\right)\Gamma_{j}^{b \;\left(N_0, N\right)}\left(k_2\right)\Gamma_{ k}^{c \;\left(N_0, N\right)}\left(k_3\right)B^{ijk}\left(N_0, k_1, k_2, k_3\right) \nonumber \\
        &+ \Gamma_{i j}^{a \;\left(N_0, N\right)}\left(k_1, k_2, k_3\right)\Gamma_{ k}^{b \;\left(N_0, N\right)}\left(k_2\right)\Gamma_{ l}^{c \; \left(N_0, N\right)}\left(k_3\right)\Sigma^{ik}\left(N_0, k_2\right)\Sigma^{jl}\left(N_0, k_3 \right) 
        \nonumber \\
        &+\Gamma_{i j}^{b \; \left(N_0, N\right)}\left(k_2, k_1, k_3\right)\Gamma_{k}^{a \;\left(N_0, N\right)}\left(k_1\right)\Gamma_{ l}^{c \; \left(N_0, N\right)}\left(k_3\right)\Sigma^{ki}\left(N_0, k_1\right)\Sigma^{jl}\left(N_0,k_3 \right)  \nonumber
        \\
        & + \Gamma_{i j}^{c \;\left(N_0, N\right)}\left(k_3, k_1, k_2\right)\Gamma_{k}^{a \;\left(N_0, N\right)}\left(k_1\right)\Gamma_{l}^{b \;\left(N_0, N\right)}\left(k_2\right)\Sigma^{ki}\left(N_0, k_1\right)\Sigma^{lj}\left(N_0, k_2 \right) \;,
    \end{align}
\end{subequations}
where the correlators at $N_0$ are obtained by using the same initial conditions discussed in Sec.~\ref{sec: hamilton to transport}. 
These can then be substituted in Eqs.~\eqref{eq: gauge transformations for Pz and Bz} to obtain $P_\zeta$ and $B_\zeta$. 
By following the procedure outlined above, one can implement numerically the MPP approach to primordial correlators. 

The traditional implementation of the transport approach described in Secs.~\ref{sec: hamilton to transport} and~\ref{sec: define correlators} and the new one with MPPs are equivalent. 
Indeed, by differentiating  Eqs.~\eqref{eq: use MPP for sigma and B at late times} and substituting Eqs.~\eqref{eq: transport eqs for MPPs} one recovers Eqs.~\eqref{eq: traditional transport equations}. 
Despite the similarities between the two approaches, there are some differences.
For example, MPPs are highly oscillatory objects, which could lead to integrator failure if a very high precision is required. 
However, this is generally not an issue for most models, as we will see, as MPPs handle high precision more effectively than \texttt{PyTransport}. 
Nonetheless, we identified a specific model where this issue becomes significant, see the discussion in footnote~\ref{foot: two-field with curved field space}. 
In contrast, \texttt{PyTransport} evolves directly phase-space correlators, which do not oscillate on sub-horizon scales. 

\section{MPPs at work}
\label{sec: MPP at work}
Since traditional numerical implementations of the transport approach have already been extensively tested against other tools to compute primordial correlators~\cite{Dias:2016rjq,Ronayne:2017qzn}, it is sufficient to test the MPP approach against results obtained with \texttt{PyTransport}. 
Not only this allows us to test the accuracy and precision of the MPP approach, but also provides insight into the comparison between these two implementations of the transport formalism.  

We devote each one of the next six sections to a different inflationary model. 
In Sec.~\ref{sec: double quadratic} we study double inflation, a two-field model with flat field-space metric and a sum-separable, quadratic potential. 
In Sec.~\ref{sec: axion-quartic} we consider again a two-field model with flat field space, but with a more complex potential form, inspired by N-flation. 
In Sec.~\ref{sec: single field with feature} we study a canonical, single-field model where the potential displays a localised feature. 
We close our tour in Sec.~\ref{sec: two field non-canonical}, where we consider a two-field model with hyperbolic field space. 

Then, in Secs.~\ref{sec: USR and large-scale bispectrum} and~\ref{sec: USR and squeezed bispectrum} we consider two single-field models featuring an ultra-slow-roll transient phase. 
The relevance of these models lies in the enhanced fluctuations that are produced on small scales, able to generate large-amplitude induced gravitational waves (see e.g. Ref.~\cite{LISACosmologyWorkingGroup:2024hsc}) and potentially form primordial black holes~\cite{Ozsoy:2023ryl}. 
These models allow us to showcase two essential applications of the MPP approach. 

\subsection{Double inflation}
\label{sec: double quadratic}

In this Section we test the MPP approach by using a model featuring two minimally coupled massive fields, with potential
\begin{equation}
    V(\phi,\, \chi) = \frac{1}{2}{m_{\phi}}^2 \, \phi^2 + \frac{1}{2}{m_{\chi}}^2 \, \chi^2 \;.
    \label{eq: double quadratic potential}
\end{equation}
This model, which goes by the name of double inflation~\cite{Silk:1986vc}, produces adiabatic and isocurvature perturbations~\cite{Polarski:1992dq, Polarski:1994rz}, which can be coupled~\cite{Langlois:1999dw}. 
Moreover, thanks to the potential~\eqref{eq: double quadratic potential} being sum-separable, it allows at least some analytical treatment within the slow-roll approximation, see e.g. Ref.~\cite{Vernizzi:2006ve} where non-Gaussianities are studied.  
Double inflation is possibly the simplest multi-field model one could consider, and it therefore provides a useful test for our algorithm without introducing additional complexities.
We set the parameters in Eq.~\eqref{eq: double quadratic potential} to $\{m_{\phi} = 1.702050 \times 10^{-6},\, m_{\chi} = 9 \, m_{\phi} \, \}$, and choose initial conditions $\{\phi_\text{in} = 12 \, , \, \chi_\text{in}=12 \, \}$, and initial velocities satisfying the slow-roll approximation. 
Here and in the following sections, all the numerical values are expressed in units in which the Planck mass is $M_{Pl} = 1$.
We utilise the double inflation model to showcase the series of tests we will perform for each model in Secs.~\ref{sec: double quadratic}-~\ref{sec: two field non-canonical}. 

First, we check the accuracy of the MPP method by comparing its output for $\mathcal{P}_\zeta(k)$ and $f_\text{NL}(k,k,k)$ in the equilateral configuration against \texttt{PyTransport}. 
\begin{figure}
    \centering
\includegraphics[width=0.5\linewidth]{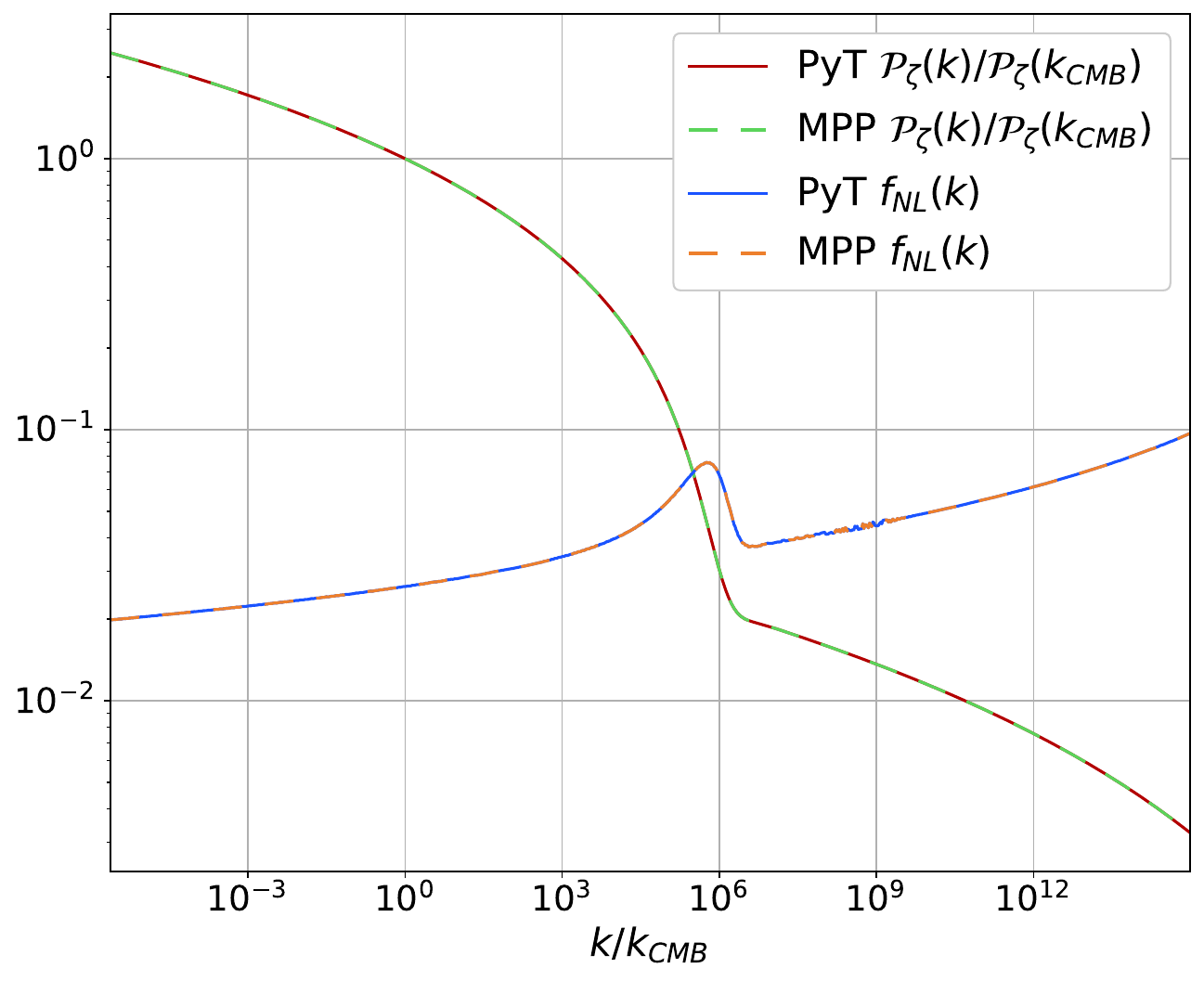}
    \caption{Numerical results for $\mathcal{P}_{\zeta}\left(k\right)/\mathcal{P}_{\zeta}\left(k_\text{CMB}\right)$ and $f_\text{NL}\left(k,k,k\right)$ obtained with \texttt{PyTransport} and MPP, for the double inflation model~\eqref{eq: double quadratic potential}. 
    We implement the corresponding initial conditions $\text{NB} = 6$ e-folds before horizon crossing, and then evolve them up to the end of inflation.
    We set the tolerances $\varepsilon=-12$ for $\mathcal{P}_\zeta$ obtained with \texttt{PyTransport}, and $\varepsilon = -13$ for the other quantities. 
    For a discussion of NB and tolerance settings see above Eq.~\eqref{eq: relative error}.
    The scales considered exited the horizon in the last $60$ e-folds of evolution.}
    \label{fig: DQ spectra}
\end{figure}
These results are represented in Fig.~\ref{fig: DQ spectra}, which shows excellent agreement between the two methods. 

We now focus on a single mode, and compare the time-dependence of phase-space 2- and 3-point correlators, see Eqs.~\eqref{eq: phase-space correlator 2pt} and~\eqref{eq: phase-space correlator 3pt}, obtained with MPPs and \texttt{PyTransport}.
We choose to work with the CMB pivot scale ($k_\text{CMB}=0.05\,\text{Mpc}^{-1}$).
Since we aim at comparing two numerical approaches (rather than at accurately computing inflationary predictions by taking into account, e.g., the effect of reheating~\cite{Martin:2014nya}), we arbitrarily associate the CMB scale with the one that crossed the horizon $50$ e-folds before the end of inflation.  
\begin{figure}
    \centering
    \includegraphics[width=0.9\linewidth]{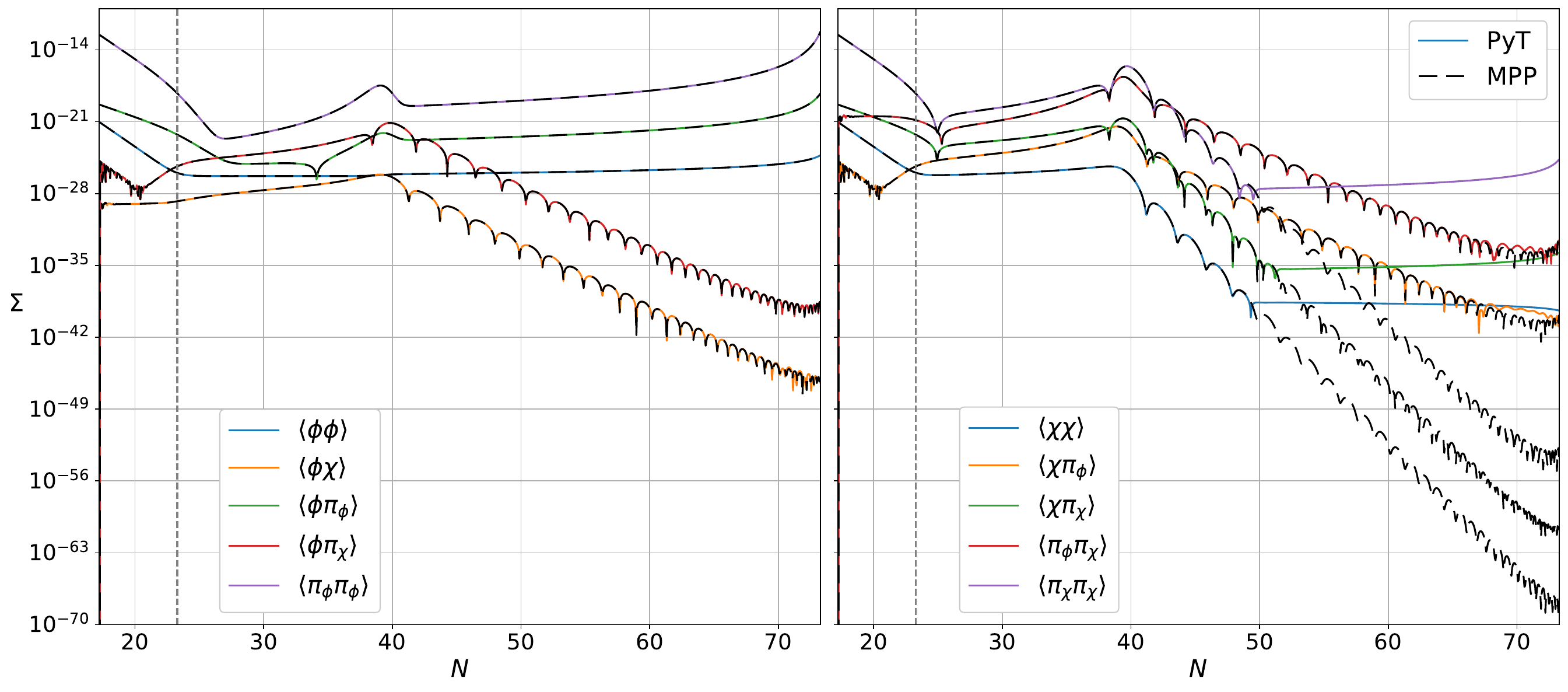}
    \caption{Time-evolution of the phase-space 2-point correlation function~\eqref{eq: phase-space correlator 2pt} of the CMB pivot scale for the double inflation model~\eqref{eq: double quadratic potential}, computed with \texttt{PyTransport} (colored) and the MPP approach (black dashed). 
    The vertical, dashed line signals the time of horizon-crossing. 
    We use $\text{NB} = 6$ e-folds of sub-horizon evolution, and tolerances $\varepsilon = -12$ for \texttt{PyTransport}, and $\varepsilon = -13$ for MPP. 
    \textit{Left panel:} 2-point correlators for which both methods give the same results. \textit{Right panel:} 2-point correlators for which the methods give different final values.}
    \label{fig: DQ confront 2pt}
\end{figure}
\begin{figure}
    \centering
    \includegraphics[width=0.9\linewidth]{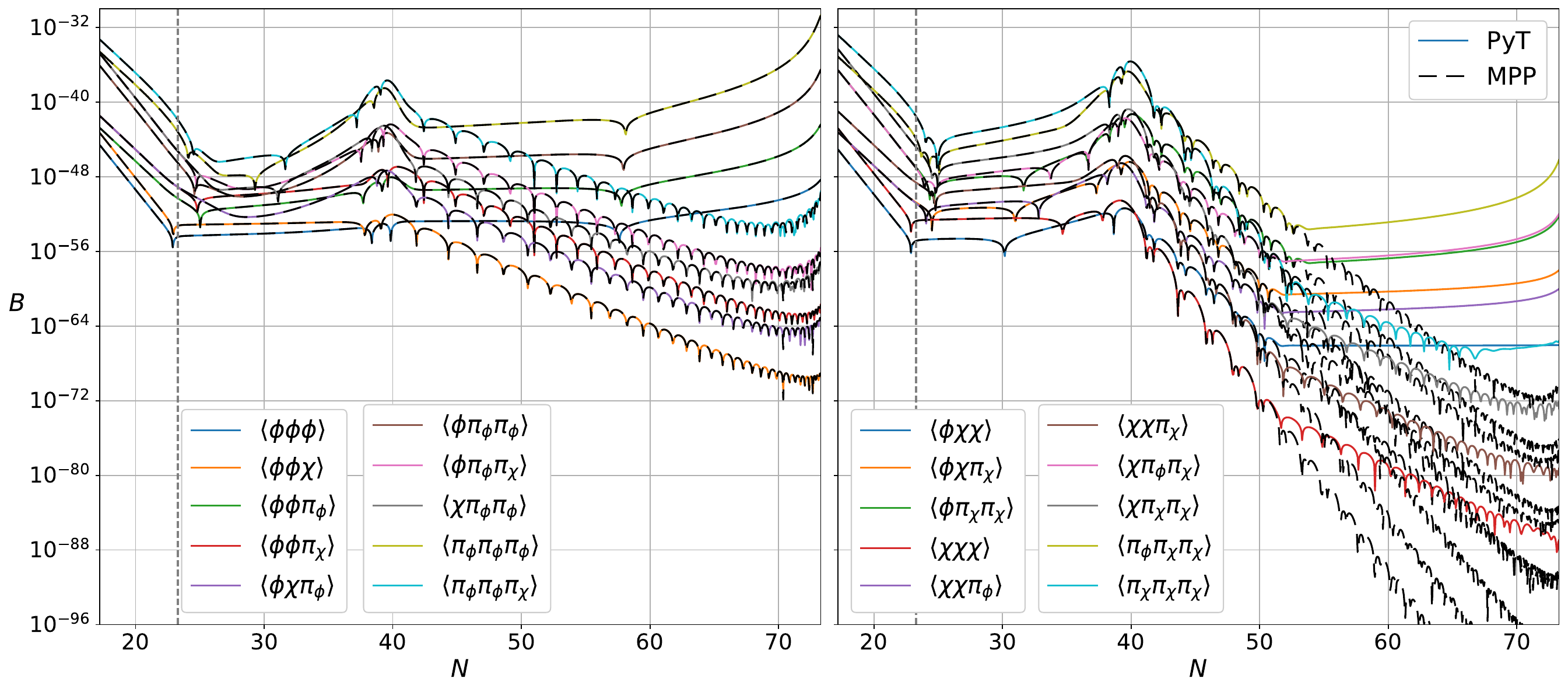}
    \caption{Time-evolution of the phase-space 3-point correlation function~\eqref{eq: phase-space correlator 3pt} of the CMB pivot scale for the double inflation model~\eqref{eq: double quadratic potential}, computed with \texttt{PyTransport} (colored) and the MPP approach (black dashed). 
    The vertical, dashed line signals the time of horizon-crossing. 
    We use $\text{NB} = 6$ e-folds of sub-horizon evolution, and tolerances $\varepsilon = -13$ for \texttt{PyTransport}, and $\varepsilon = -13$ for MPP. 
    \textit{Left panel:} 3-point correlators for which both methods give the same results. \textit{Right panel:} 3-point correlators for which the methods produce different final values.}
    \label{fig: DQ confront 3pt}
\end{figure}

We represent the results for the 2- and 3-point correlation functions in Figs.~\ref{fig: DQ confront 2pt} and~\ref{fig: DQ confront 3pt}. 
Here, and in the next sections, we label the field perturbations with the name of the field (e.g. $Q^\phi \longrightarrow \phi$) and the conjugate momentum perturbation with the letter $\pi$ (e.g. $P_\phi \longrightarrow \pi_\phi$).

In the left panel of Fig.~\ref{fig: DQ confront 2pt}, we show that the results for the three correlators with largest magnitude at the end of inflation are in agreement between the two methods. 
On the other hand, the super-horizon decay of the correlators in the right panel, which are those of a heavy field, is precisely tracked by MPPs, but not by \texttt{PyTransport}. 
We observe the same behavior in the 3-point correlators shown in Fig~\ref{fig: DQ confront 3pt}.
We note that these results are compatible with those in Fig.~\ref{fig: DQ spectra}, as both MPPs and \texttt{PyTransport} yield a very small final value (albeit different) for the right-panels correlators, while they agree on those with largest magnitude. 
In other words, in this case the inability of \texttt{PyTransport} to precisely track decaying correlators does not affect the final result, but one can imagine situations in which this won't be the case, e.g. for the models we consider in Secs.~\ref{sec: USR and large-scale bispectrum} and~\ref{sec: USR and squeezed bispectrum}. 

We then investigate the performance of MPPs in terms of (i) the number of sub-horizon e-folds of evolution, NB, and (ii) tolerance settings. 
In both approaches, these parameters are set by the user. 
NB corresponds to the number of e-folds before horizon crossing when the initial conditions for phase-space correlators (\texttt{PyTransport}) and MPPs (MPP approach) are imposed. 
For \texttt{PyTransport}, by assuming that the initial conditions are set sufficiently far inside of the horizon, the phase-space variables behave like massless and non-interacting fields in de Sitter, and their initial conditions can be computed accordingly~\cite{Dias:2015rca, Ronayne:2017qzn}.
On one hand the parameter NB must be sufficiently large to ensure this description is correct, on the other the longer the modes are evolved on sub-horizon the slower the computation will be, if possible at all. 
We test values $\text{NB}\in [1, \, 14]$. 
Tolerances are parameters in an ODE solver that determine the precision of its output. 
The relative tolerance establishes a threshold for the acceptable relative error at each time step, while the absolute tolerance defines the allowable absolute difference between consecutive time steps.
We set relative and absolute tolerance to the same value and parametrize them in terms of $\varepsilon \equiv \log_{10}{\left(\text{tol}\right)}$, which we vary in the interval $\varepsilon \in [-18, \, -5]$. 

To perform our tests, we compute the values at the end of inflation of the dimensionless power spectrum at $k_\text{CMB}$, and the amplitude of the equilateral non-Gaussianity evaluated at the same scale, \textit{i.e.} $k_1=k_2=k_3=k_\text{CMB}$ in Eq.~\eqref{eq: fnl}.
For each value of NB, we first find the largest tolerance which allows to attain a specific level of precision in $\mathcal{P}_\zeta(k_\text{CMB})$ and $f_\text{NL}(k_\text{CMB})$.
We start by fixing $\varepsilon$ to its largest value, and compute the corresponding $\mathcal{P}_\zeta(k_\text{CMB})$ and $f_\text{NL}(k_\text{CMB})$. 
We then repeat the computation by decreasing the tolerance $\varepsilon$ in steps of $\Delta \varepsilon=0.5$, until $\varepsilon=-18$ is reached. 
We define the relative error between one step and the following one as 
\begin{equation}
    \Delta Q = \left|\frac{Q^{(n-1)} - Q^{(n)}}{Q^{(n)}}\right| \;, 
    \label{eq: relative error}
\end{equation}
where $Q$ is a place-holder for $\mathcal{P}_{\zeta}(k_\text{CMB})$ or $f_\text{NL}(k_\text{CMB})$, and the index $n$ enumerates the steps.
The quantity $\Delta Q$ measures the relative change in $Q$ between computations performed with different tolerances, and as $\varepsilon$ decreases we expect $\Delta  Q$ to  decrease overall. 
We arbitrarily set the precision threshold to be $\theta=10^{-8}$, and define the optimal tolerance, $\tilde \varepsilon$, to be the value of $\varepsilon$ for which $\Delta Q$ stably drops below $\theta$, signaling that the numerical solution has converged at the desired level of precision.
This allows us to find, for fixed NB, the optimal operating set-up in terms of relative and absolute tolerances for each code (see Ref.~\cite{Mulryne:2016mzv}) .
We repeat this procedure for all values of NB.

\begin{figure}
    \centering
    \includegraphics[width=0.9\linewidth]{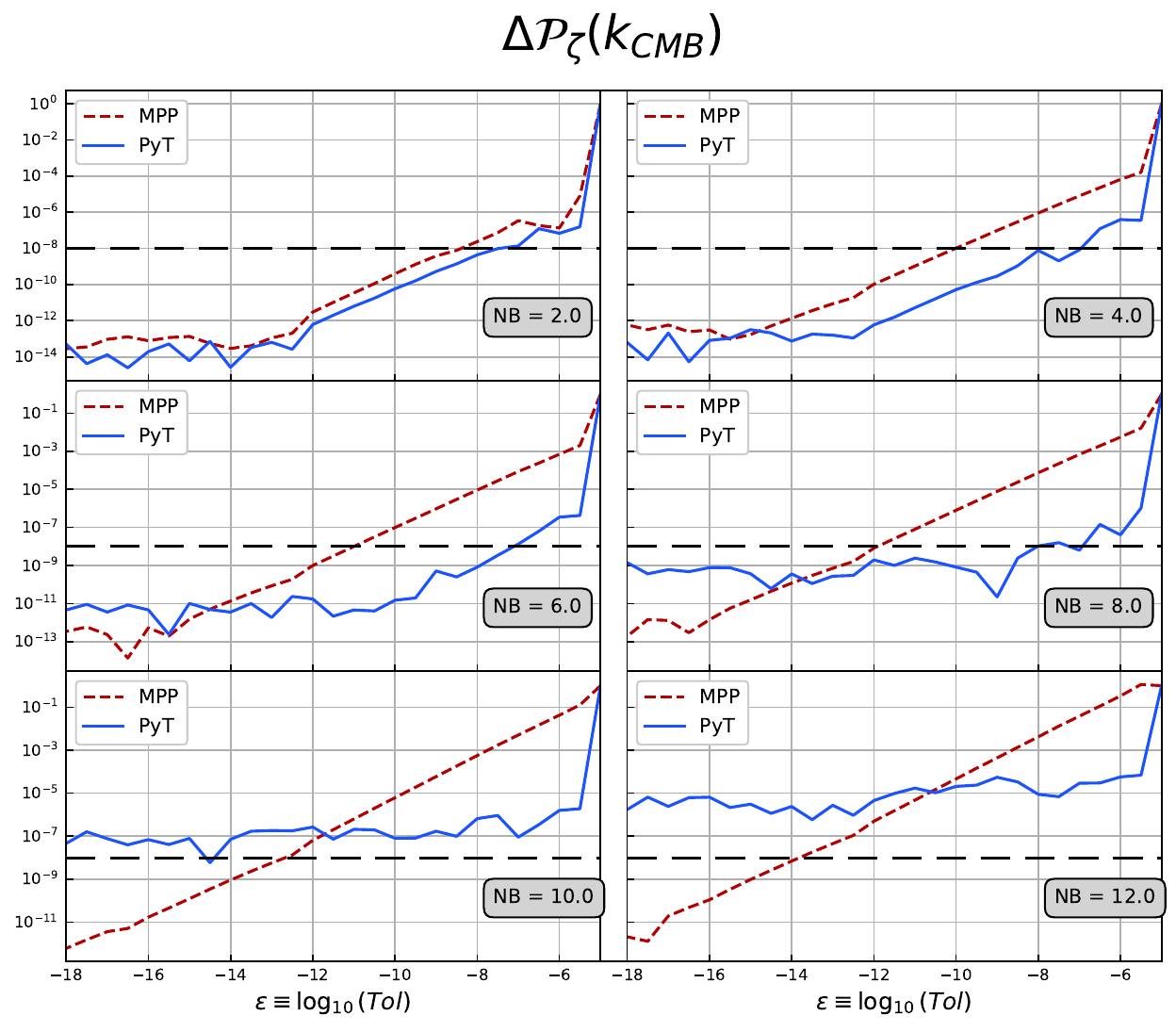}
    \caption{Relative error in the computation of the CMB mode power spectrum, $\Delta \mathcal{P}_\zeta(k_\text{CMB})$, see Eq.~\eqref{eq: relative error}, as a function of the tolerance, $\varepsilon \equiv \log_{10}(\text{tol})$.  
    Each panel corresponds to a different value of the number of e-folds of sub-horizon evolution, NB.  
    The horizontal line in each panel represents the accuracy threshold $\theta = 10^{-8}$, which we use to define the optimal tolerance $\tilde \varepsilon$. } 
    \label{fig: DQ 2pt NB test}
\end{figure}
The numerical results for $\Delta \mathcal{P}_\zeta(k_\text{CMB})$ are represented in Fig.~\ref{fig: DQ 2pt NB test} for six values of NB.
As expected, for both codes $\Delta \mathcal{P}_\zeta(k_\text{CMB})$ overall decreases for smaller $\varepsilon$ values. 
For the first four NB values, \texttt{PyTransport} reaches the threshold before MPP, i.e. at larger tolerance. 
However, for $\text{NB} > 9.5$ MPP crosses the threshold consistently, while for \texttt{PyTransport} $\Delta \mathcal{P}_\zeta(k_\text{CMB})$ saturates and the code is not able to achieve the desired precision.  
For this model, we therefore conclude that with MPP the precision consistently gets better for smaller $\varepsilon$, even when NB is large. 

\begin{figure}
    \centering
    \includegraphics[width=0.9\linewidth]{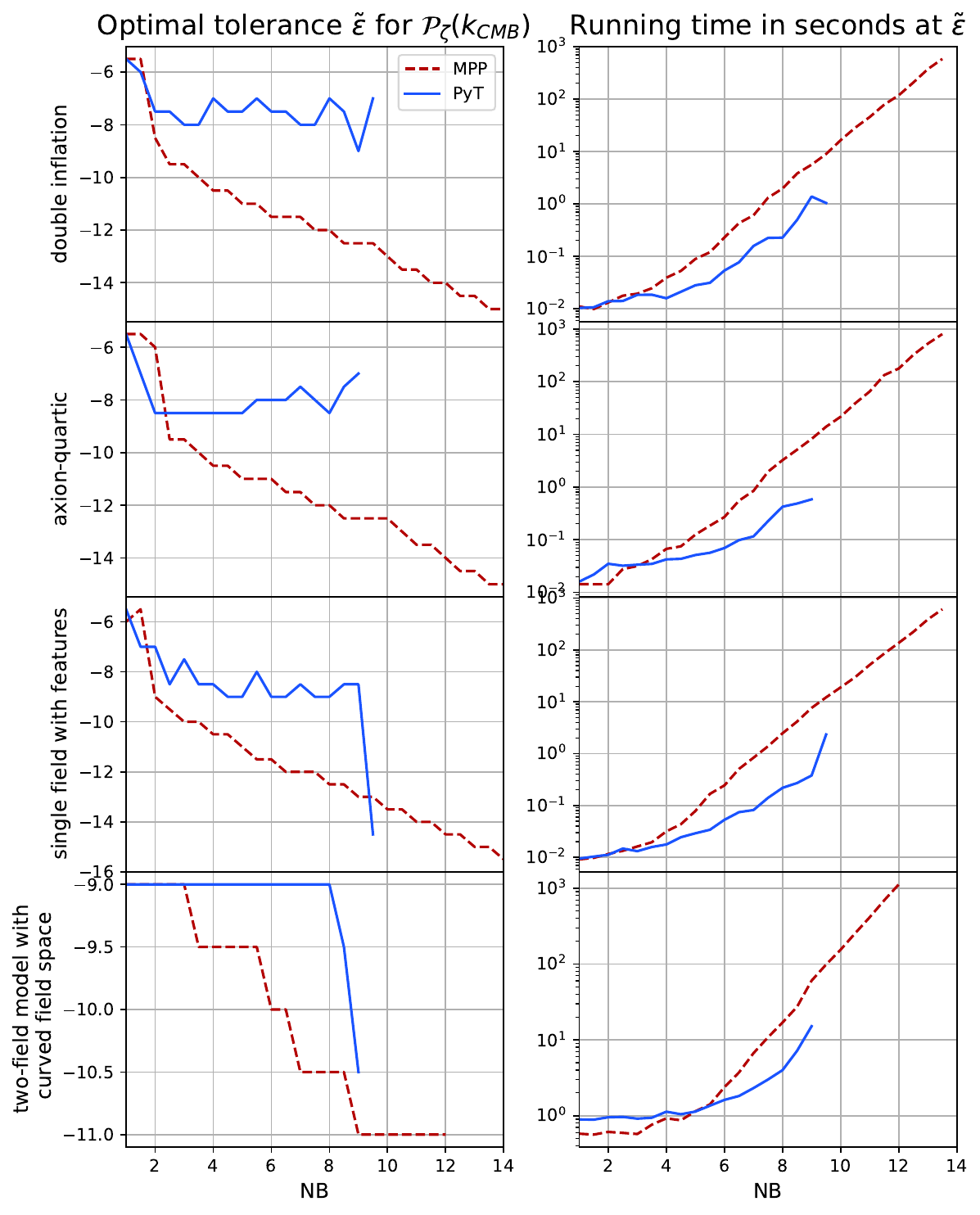}
    \caption{\textit{Left panel:} optimal tolerance $\tilde \varepsilon$ for computing $\mathcal{P}_\zeta(k_\text{CMB})$ as a function NB. 
    \textit{Right panel:} running time in seconds as a function of NB for a run performed with optimal tolerance $\tilde \varepsilon$.
    Each row corresponds to one of the models discussed in Secs.~\ref{sec: double quadratic}-~\ref{sec: two field non-canonical}, as detailed in the left-hand-side label.}
    \label{fig: summary plot 2pt}
\end{figure}
In the first row, left panel of Fig.~\ref{fig: summary plot 2pt} we summarize the results found in Fig.~\ref{fig: DQ 2pt NB test} for the optimal tolerance, $\tilde \varepsilon$, as a function of NB.
In the right panel of the same row we represent the running time for computing $\mathcal{P}_\zeta(k_\text{CMB})$ with different NB values and the corresponding value of the optimal tolerance $\tilde \varepsilon$. 
When \texttt{PyTransport} is able to achieve the desired precision, it does so faster than MPP. 
Nevertheless, if we extrapolate the scaling of the running time with respect to NB, in comparison the MPP algorithm performs better for larger NB, and we recognise the same behavior across all models tested.  

\begin{figure}
    \centering
    \includegraphics[width=0.9\linewidth]{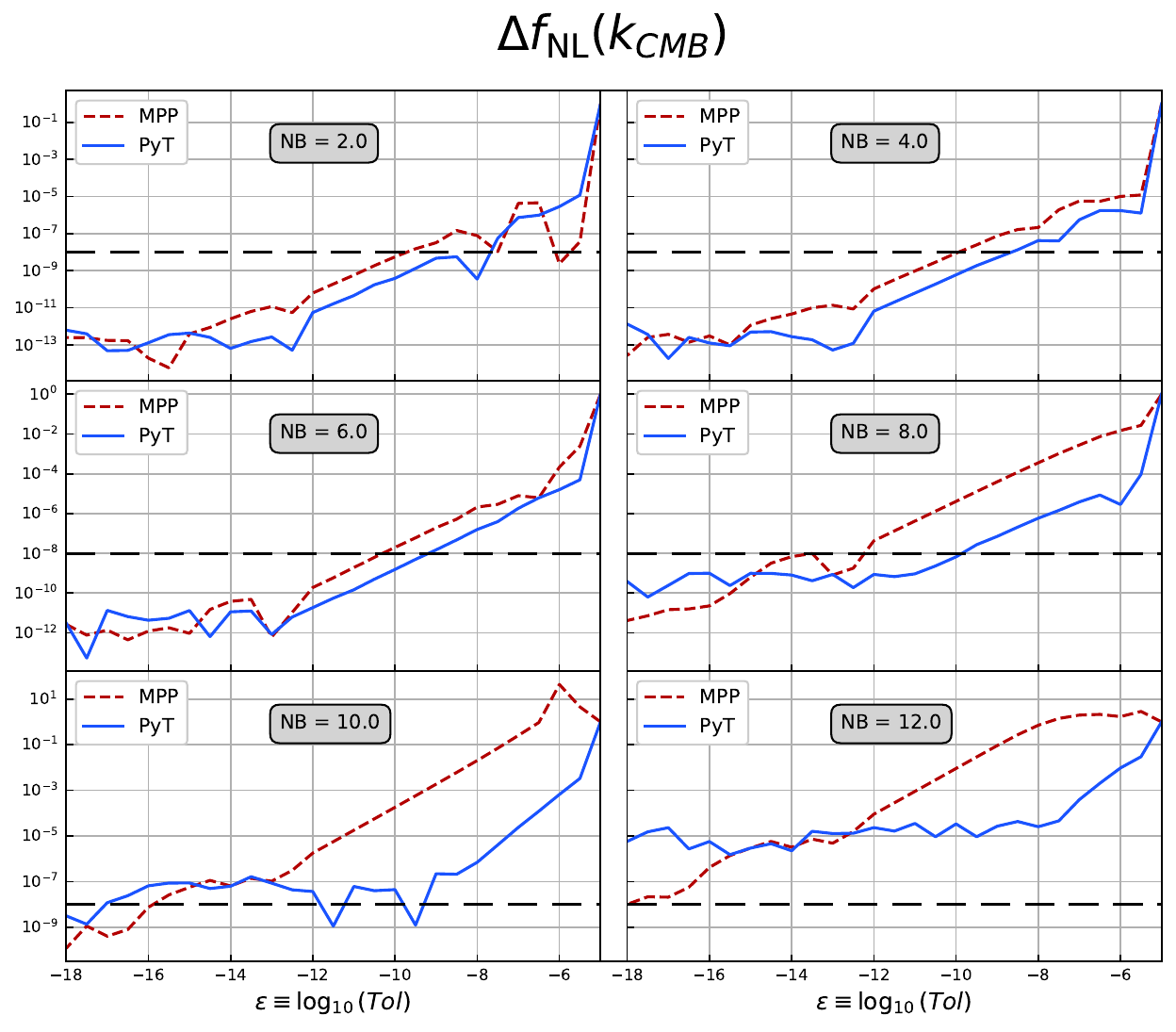}
    \caption{Relative error in the computation of the equilateral non-Gaussianity at the CMB scale, $\Delta f_\text{NL}(k_\text{CMB})$, see Eq.~\eqref{eq: relative error}, as a function of the tolerance, $\varepsilon \equiv \log_{10}(\text{tol})$.  
    Each panel corresponds to a different value of the number of e-folds of sub-horizon evolution, NB.  
    The horizontal line in each panel represents the accuracy threshold $\theta = 10^{-8}$, which we use to define the optimal tolerance $\tilde \varepsilon$.}
    \label{fig: DQ 3pt NB tests}
\end{figure}
\begin{figure}
    \centering
    \includegraphics[width=0.9\linewidth]{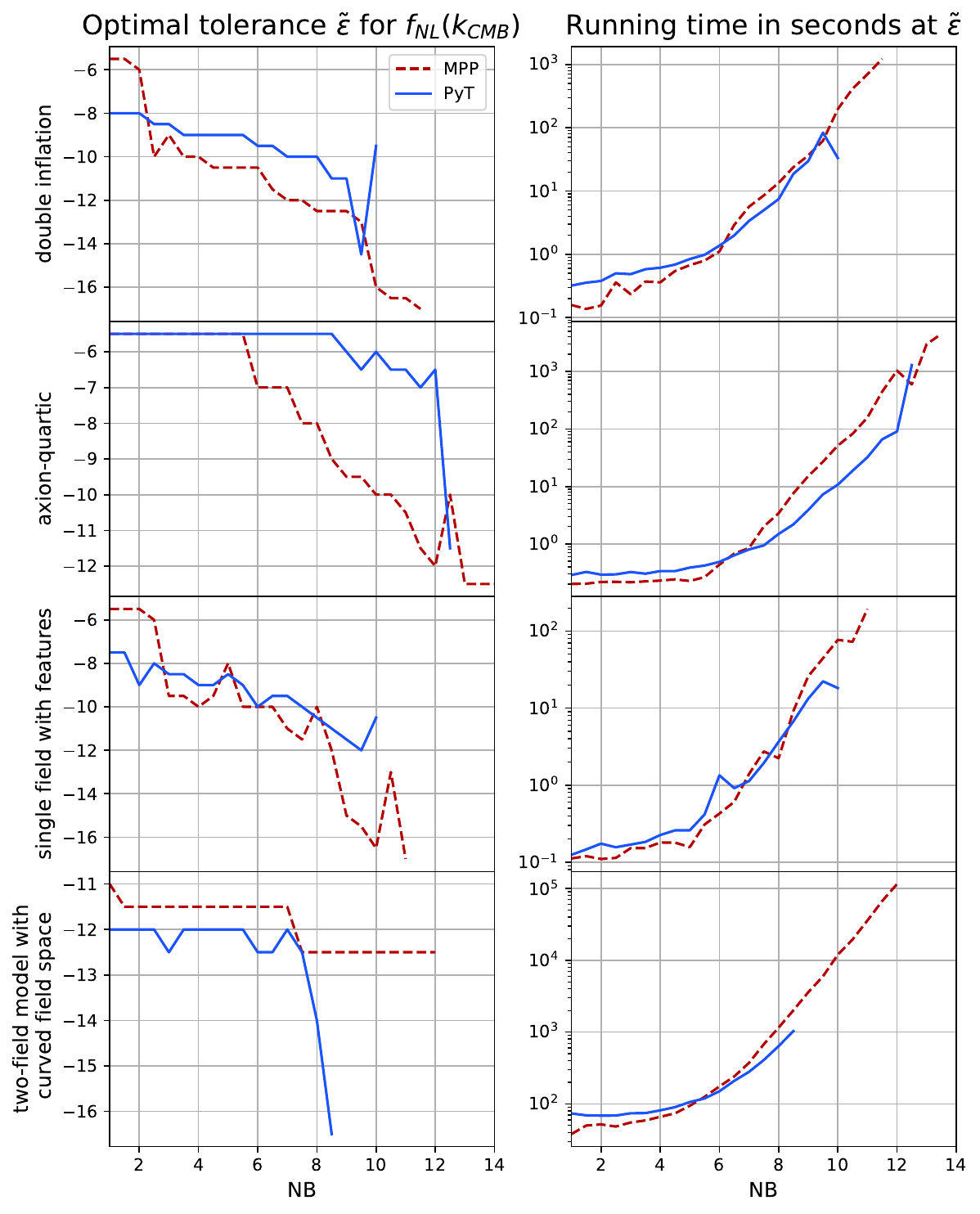}
    \caption{\textit{Left panel:} optimal tolerance $\tilde \varepsilon$ for computing $f_\text{NL}(k_\text{CMB})$ as a function NB. 
    \textit{Right panel:} running time in seconds as a function of NB for a run performed with optimal tolerance $\tilde \varepsilon$.
    Each row corresponds to one of the models discussed in Secs.~\ref{sec: double quadratic}-~\ref{sec: two field non-canonical}, as detailed in the left-hand-side label.}
    \label{fig: summary plot 3pt}
\end{figure}
Fig.~\ref{fig: DQ 3pt NB tests} and the first row in Fig.~\ref{fig: summary plot 3pt} show our findings for the case of $f_\text{NL}(k_\text{CMB})$. 
For low NB, \texttt{PyTransport} can reach the threshold for larger tolerance than MPP, but does not reach the desired precision for $\text{NB} > 10$. 
We can reach the threshold with MPP up to $\text{NB}=11.5$. 
By inspecting the summary plots in Fig.~\ref{fig: summary plot 3pt}, we see a sightly different behavior of the running time with respect to the case of $\mathcal{P}_\zeta(k_\text{CMB})$. 
The running times of the two codes are now comparable, and seem to scale with NB in the same way. 

The numerical results discussed in this section show that for double inflation the MPP approach produces accurate results, which are more precise than those of \texttt{PyTransport} in describing decaying correlators. 
In terms of performance at optimal tolerance, we have found that MPP are either slightly slower or comparable to \texttt{PyTransport}, but (if needed) they allow to reach high precision even when NB is large. 

\subsection{Axion-quartic}
\label{sec: axion-quartic}

The second model we consider is a two-field model with flat field-space metric and potential 
\begin{equation}
    V(\phi,\, \chi) = \frac{1}{4}
    g\, \phi^4 + V_0 \left[1 - 
    \cos{\left( \frac{2\pi\chi}{f}\right)}\right]\;.
    \label{eq: axion quartic potential}
\end{equation}
The model is inspired by N-flation~\cite{Dimopoulos:2005ac} and it was introduced in Ref.~\cite{Elliston:2012wm}. 
In N-flation the potential landscape is given by adding the single contributions of many axion fields, which cooperatively support inflation~\cite{Dimopoulos:2005ac}.
In Refs.~\cite{Kim:2010ud, Kim:2011jea} it was shown that the model's predictions are dominated by the field that starts the closest to the hilltop of its potential, and therefore Ref.~\cite{Elliston:2011dr} proposed an effective version of the model, featuring the dominant axion, the $\chi$ field in Eq.~\eqref{eq: axion quartic potential}, while the collective effect of the remaining axions is represented by an effective field $\phi$, with potential $m^2\phi^2$. 
In Ref.~\cite{Elliston:2012wm} the $\phi$ potential was chosen instead to be quartic, in order to provide large-scale predictions in closer agreement with observations. 
The model~\eqref{eq: axion quartic potential} produces non-Gaussianity peaked in the squeezed configuration and with amplitude $f_\text{NL}\sim \mathcal{O}(10)$~\cite{Dias:2016rjq}. 
 
In our numerical simulations we fix the potential parameters $f = 1$, and $V_0 = (25)^2/(2\pi)^2 g$, while $g$ is chosen such that the amplitude of the power spectrum is compatible with \textit{Planck} measurements. 
We set the initial conditions $\{\phi_\text{in} = 23.5\, , \, \chi_\text{in} = f/2 - 10^{-3} \,\}$, and initial velocities equal to zero. 

We compute the relative errors $\Delta \mathcal{P}_\zeta(k_\text{CMB})$ and $\Delta f_\text{NL}(k_\text{CMB})$ numerically, following the methodology described in Sec.~\ref{sec: double quadratic}, and summarise the results in the second row of Figs.~\ref{fig: summary plot 2pt} and~\ref{fig: summary plot 3pt} respectively. 
In the left panel of Fig.~\ref{fig: summary plot 2pt} we observe that within the MPP approach the optimal tolerance can be defined across all NB values, whereas with \texttt{PyTransport} computations are limited to $\text{NB} \leq 9$. 
Within the range of NB in which both codes reach the desired precision, \texttt{PyTransport} requires larger $\tilde \varepsilon$ than MPP. 
In terms of computational time, \texttt{PyTransport} is faster for $\text{NB} \in [3,9]$, while for very low NB values, MPP outperforms \texttt{PyTransport}. 

A similar trend is observed in the analysis of $\Delta f_\text{NL}(k_\text{CMB})$, see Fig.~\ref{fig: summary plot 3pt}. 
At lower NB values, \texttt{PyTransport} reaches the target precision at a higher tolerance, while MPP remains effective over a wider NB range. The right panel illustrates that MPP is computationally faster for $\text{NB} \leq 6$, beyond which \texttt{PyTransport} becomes more efficient. 

\subsection{Single-field model with feature}
\label{sec: single field with feature}

We consider a canonical, single-field model, where the inflaton potential displays a localised feature at $\phi=\phi_0$~\cite{Adams:2001vc}
\begin{equation}
    V(\phi) = \frac{1}{2} m^2 \phi^2 \left[1 + c\tanh{\left(\frac{\phi - \phi_0}{d}\right)}\right] \;. 
    \label{eq: single field potential}
\end{equation}
For initial conditions $\phi_\text{in}>\phi_0$, the inflaton rolls down its quadratic potential and crosses a step at $\phi_0$, which is downward (upward) for $c>0$ ($c<0$) and whose sharpness depends on the value of $d$. 
The presence of a sharp feature can induce interesting dynamics, for example significant deviation from slow-roll and scale-dependent oscillations in the power spectrum~\cite{Adams:2001vc} and bispectrum~\cite{Chen:2006xjb, Chen:2008wn}. 
For these reasons, the single-field potential~\eqref{eq: single field potential} provides a good testing ground for our MPP algorithm. 
We set $\{c = 0.0018, \, d = 0.022\,, \, \phi_0 = 14.84 \, \}$, and initial condition $\phi_\text{in} = 16.5$ and velocity satisfying the slow-roll approximation. 

We compute the relative errors $\Delta \mathcal{P}_\zeta(k_\text{CMB})$ and $\Delta f_\text{NL}(k_\text{CMB})$ numerically, following the methodology described in Sec.~\ref{sec: double quadratic}, and summarise the results in the third row of Figs.~\ref{fig: summary plot 2pt} and~\ref{fig: summary plot 3pt} respectively. 
For the power spectrum case, \texttt{PyTransport} fails to reach the required precision for \(\text{NB} \geq 10\), whereas MPPs successfully reach the threshold for every NB. 
In terms of computational speed, the MPP method is slower than \texttt{PyTransport}. 
Nevertheless, the (extrapolated) scaling of the running time with respect to NB shows that the MPP algorithm performs better for larger NB. 
For the bispectrum case, we observe that MPP can reach the desired precision up to $\text{NB}=11$, one e-folds more than what \texttt{PyTransport} can achieve. 
For the NB values that both methods run with, the corresponding running times are comparable.

\subsection{Two-field model with curved field space}
\label{sec: two field non-canonical}
We consider now a two-field model of inflation with non-trivial field-space metric. 
In particular, we choose to work with the model proposed by Braglia \textit{et al} in Ref.~\cite{Braglia:2020eai}, featuring two fields, $\phi$ and $\chi$, living on a hyperbolic field space, with action 
\begin{equation}
    \mathcal{S} = \int \mathrm{d}^4x \; \sqrt{-g}\left[ \frac{R}{2}\, - \frac{1}{2}\left(\partial\phi\right)^2 - \frac{1}{2}e^{2b_1\phi}\left(\partial\chi\right)^2 - V(\phi,\chi) \right],
    \label{eq: braglia action}
\end{equation}
and potential
\begin{equation}
    V(\phi,\chi) = V_0 \, \frac{\phi^2}{\phi_0^2 + \phi^2} + \frac{1}{2}{m_{\chi}}^2 \, \chi^2 \;. 
    \label{eq: braglia potential}
\end{equation}
Geometrical effects can lead to enhanced scalar fluctuations on small scales, which could in turns produce primordial black holes and large second-order gravitational waves~\cite{Braglia:2020eai}. 
In Ref.~\cite{Iacconi:2023slv} it is shown that 
for models which potentially lead to interesting phenomenology at peak scales, the tree-level scalar power spectrum receives significant non-linear corrections, signalling the breakdown of perturbative treatments.

In the following, we set $\{\phi_0 = \sqrt{6},\, {m_{\chi}}^2 = V_0 /500, \, b_1 = 7.6\}$ and $V_0$ to the value required to match \textit{Planck} observations. 
We choose initial conditions $\{\phi_\text{in} = 7\, , \, \chi_\text{in} = 7.31 \,\}$ and initial velocities satisfying the slow-roll approximation. 

We compute\footnote{\label{foot: two-field with curved field space}For this model we found that the MPP code was not able to run for very stringent tolerances ($\varepsilon<10^{-8}$), due to the integrator failing in the first time-step of evaluation. 
To circumvent this problems, we reduced the tolerance during the initialization phase (with an initialization step $\Delta N_\text{init} = 0.01$), and then reset it to the desired value for the rest of the integration.} the relative errors $\Delta \mathcal{P}_\zeta(k_\text{CMB})$ and $\Delta f_\text{NL}(k_\text{CMB})$ numerically, following the methodology described in Sec.~\ref{sec: double quadratic}, and summarise the results in the fourth row of Figs.~\ref{fig: summary plot 2pt} and~\ref{fig: summary plot 3pt} respectively. 
Results for  $\Delta \mathcal{P}_\zeta(k_\text{CMB})$ reveal evident differences between the two computational approaches. 
For example, \texttt{PyTransport} maintains a constant optimal tolerance for \(\text{NB} \leq 8\), before dropping to \(\tilde{\varepsilon} = -10.5\) at \(\text{NB} = 9\), beyond which it fails to reach the desired precision. 
In contrast, the MPP algorithm exhibits a steady decrease in the optimal tolerance values up to \(\text{NB} = 12\), where it can no longer reach the desired precision. 
MPPs are initially faster for low \(\text{NB}\) values, while \texttt{PyTransport} becomes more efficient for \(\text{NB} > 5\).
A similar trend is observed for $\Delta f_\text{NL}(k_\text{CMB})$, where the MPP method reaches the desired precision over a broader range of \(\text{NB}\) values. 
MPPs display a more stable decreasing trend for $\tilde \varepsilon$, up to \(\text{NB} = 12\). The running times are comparable for NB values for which the comparison is possible, with \texttt{PyTransport} being marginally slower for \(\text{NB} \leq 5.5\) and slightly faster at higher \(\text{NB}\).

\subsection{Ultra-slow-roll model: large-scale bispectrum}
\label{sec: USR and large-scale bispectrum}

In the literature \cite{Atal:2018neu}, and in our own testing\footnote{We acknowledge also private communication with Pippa Cole and David Seery on this matter.}, there are documented examples where \texttt{PyTransport} leads to incorrect results. We now test the MPP algorithm on such a scenario.
We consider a single-field, canonical model, with potential 
\begin{equation}
    V(\phi) = V_0 \left[ 1 - A\tanh\left(\frac{\phi - \phi_0}{d} \right) \right] \left[ 1 - \frac{B\phi^2}{1 + \phi/f} \right] \;. 
    \label{eq: PBH potential}
\end{equation}
In the following, we adopt the parameter set
$\{V_0 = 1, \, A = 0.003, \, \phi_0 = 0.005, \, d = 0.0005, \, B = 0.06, \, f = 0.001 \}$, with initial condition $\phi_\text{init} = 0.0008$ and initial velocity determined by the slow-roll approximation.
Note that the potential~\eqref{eq: PBH potential} does not lead inflation to an end, and should therefore be regarded as a toy-model.  
We manually halt the evolution after $50$ e-folds. 

The model~\eqref{eq: PBH potential} belongs to a class of inflationary scenarios called ``inflation falls'', in which the inflationary potential becomes steeper and then shallower again,  -- i.e. the inflaton potential causes the field velocity to accelerate and then decelerate through a phase of ultra-slow-roll (USR)~\cite{Inomata:2021uqj, Carrilho:2019oqg, Hertzberg:2017dkh}. The potential above is engineered to exhibit these features. 
The Hubble slow-roll parameters are defined as
\begin{equation}
    \epsilon_1 = - \frac{\dot H}{H^2} \quad \text{and} \quad \epsilon_i = \frac{\dot \epsilon_{i-1}}{H\epsilon_{i-1}} \; \forall i\geq 2 \;, 
    \label{eq: slow-roll parameter}
\end{equation}
where $H$ is the Hubble rate during inflation and we relabel $\epsilon_2=\eta$.
Slow-roll dynamics is characterised by $\epsilon_i\ll 1\quad \forall\,  i$, while during USR $\epsilon_1\ll 1$ and $\eta \sim -6$. 
The transient ultra-slow-roll phase generates a pronounced peak in the power spectrum on small scales, leading to amplified gravitational waves induced at second-order in perturbation theory, as well as possible primordial black hole production.
\begin{figure}
    \centering
    \includegraphics[width=0.5\linewidth]{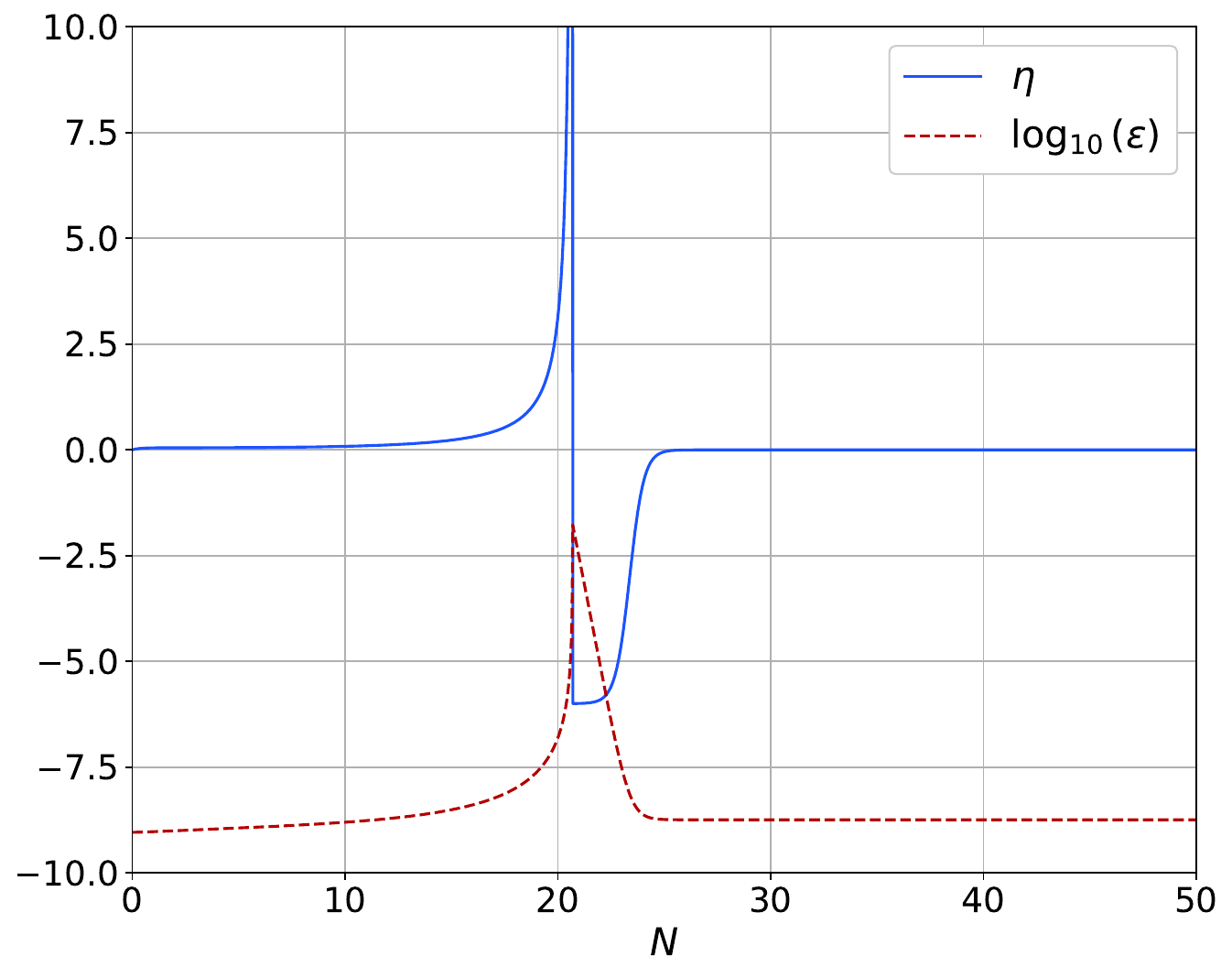}
    \caption{Time-evolution of the first two slow-roll parameters, $\epsilon$ and $\eta$, for the ultra-slow-roll model~\eqref{eq: PBH potential}.}
    \label{fig: PBH background}
\end{figure}
In Fig.~\ref{fig: PBH background} we display the time-evolution of the first two slow-roll parameters, see Eq.~\eqref{eq: slow-roll parameter}. 
Around $N=22$ the velocity grows before the inflaton enters the transient ultra-slow-roll phase $(\eta\sim -6)$.

\begin{figure}
    \centering
    \begin{subfigure}{.45\textwidth}
        \includegraphics[width=\textwidth]{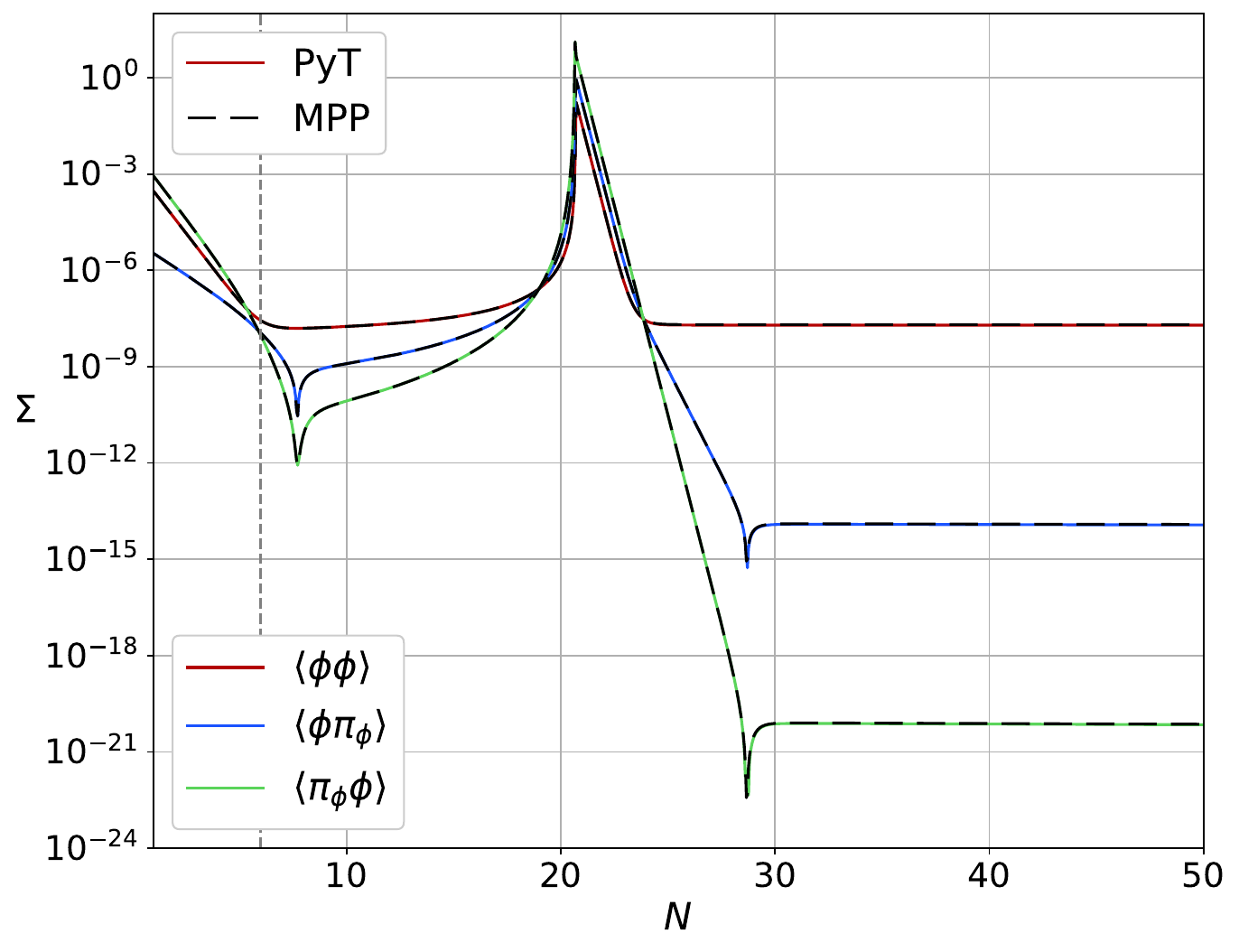}
    \end{subfigure}
    \begin{subfigure}{.45\textwidth}
        \includegraphics[width=\textwidth]{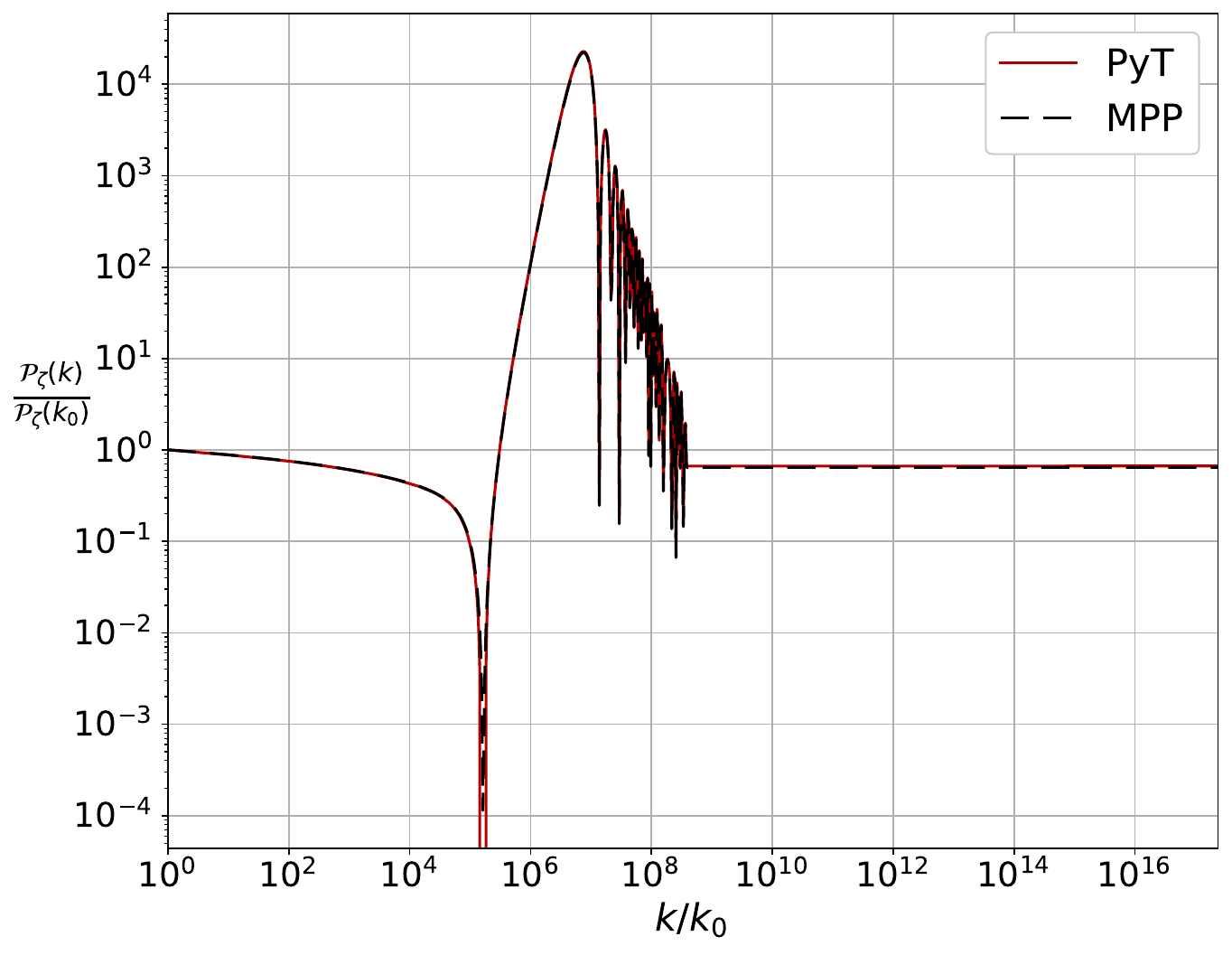}
    \end{subfigure}
    \caption{\textit{Left panel:} evolution of phase-space 2-point correlators for a large-scale mode, $k_0$, that exits the horizon at \(N = 6\).
    Continuous (dashed) lines represent \texttt{PyTransport} (MPP) results. 
    \textit{Right panel:} scalar power spectrum, $\mathcal{P}_\zeta(k)/\mathcal{P}_\zeta(k_0)$, computed with both methods.}
    \label{fig: PBH Two-point and Pz}
\end{figure}
In the left panel of Fig.~\ref{fig: PBH Two-point and Pz} we display the time-evolution of the 2-point phase-space correlators, which show excellent agreement between the results obtained with the MPP and \texttt{PyTransport} approaches. 
In the right panel $\mathcal{P}_\zeta(k)$ is displayed for scales that exit the horizon between 44 and 4 e-folds before the evolution is manually interrupted. 
Consistently with the results shown in the left panel, both methods produce the same scalar power spectrum. 

\begin{figure}
    \centering
    \begin{subfigure}{.45\textwidth}
        \includegraphics[width=\textwidth]{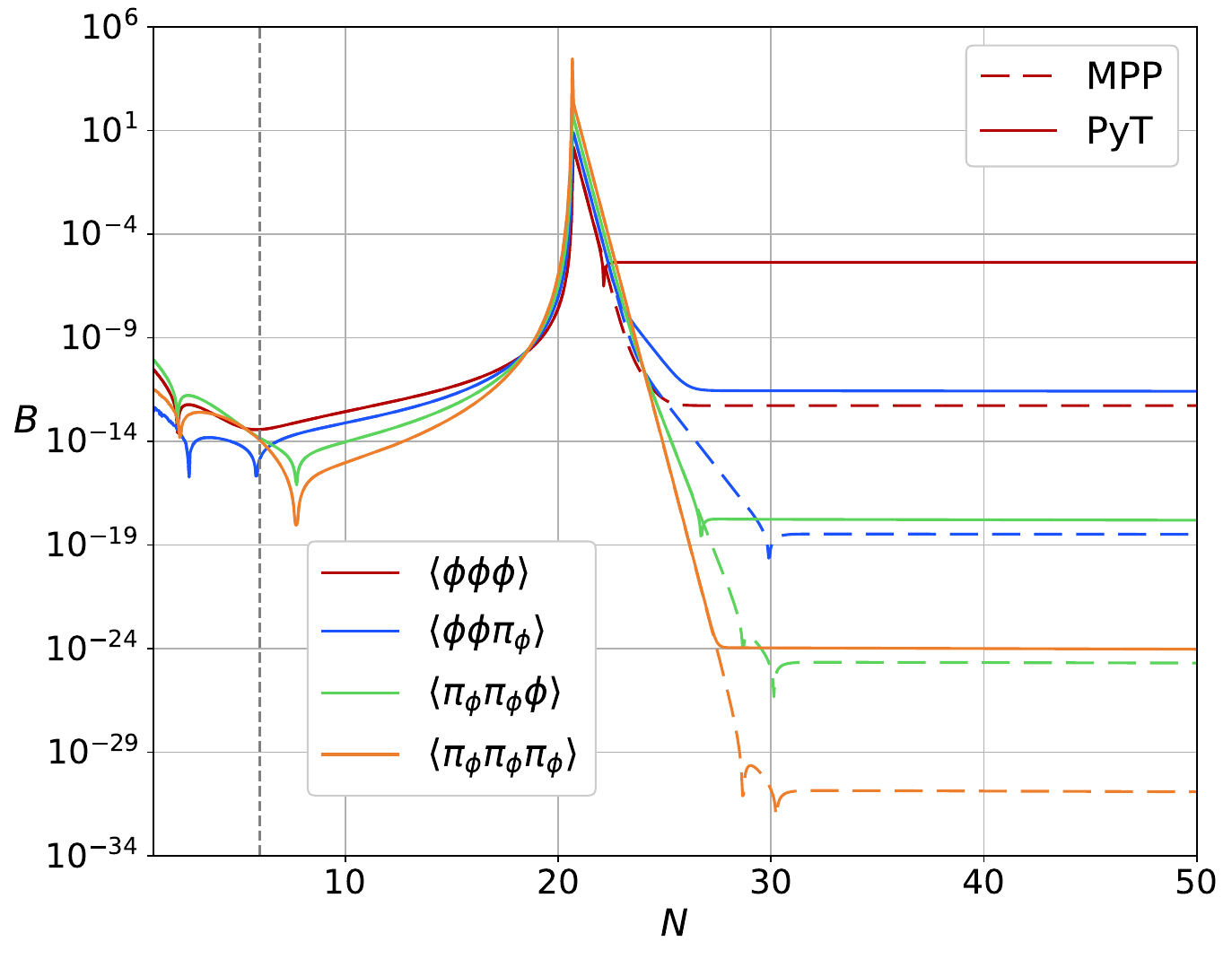}
    \end{subfigure}
    \begin{subfigure}{.45\textwidth}
        \includegraphics[width=\textwidth]{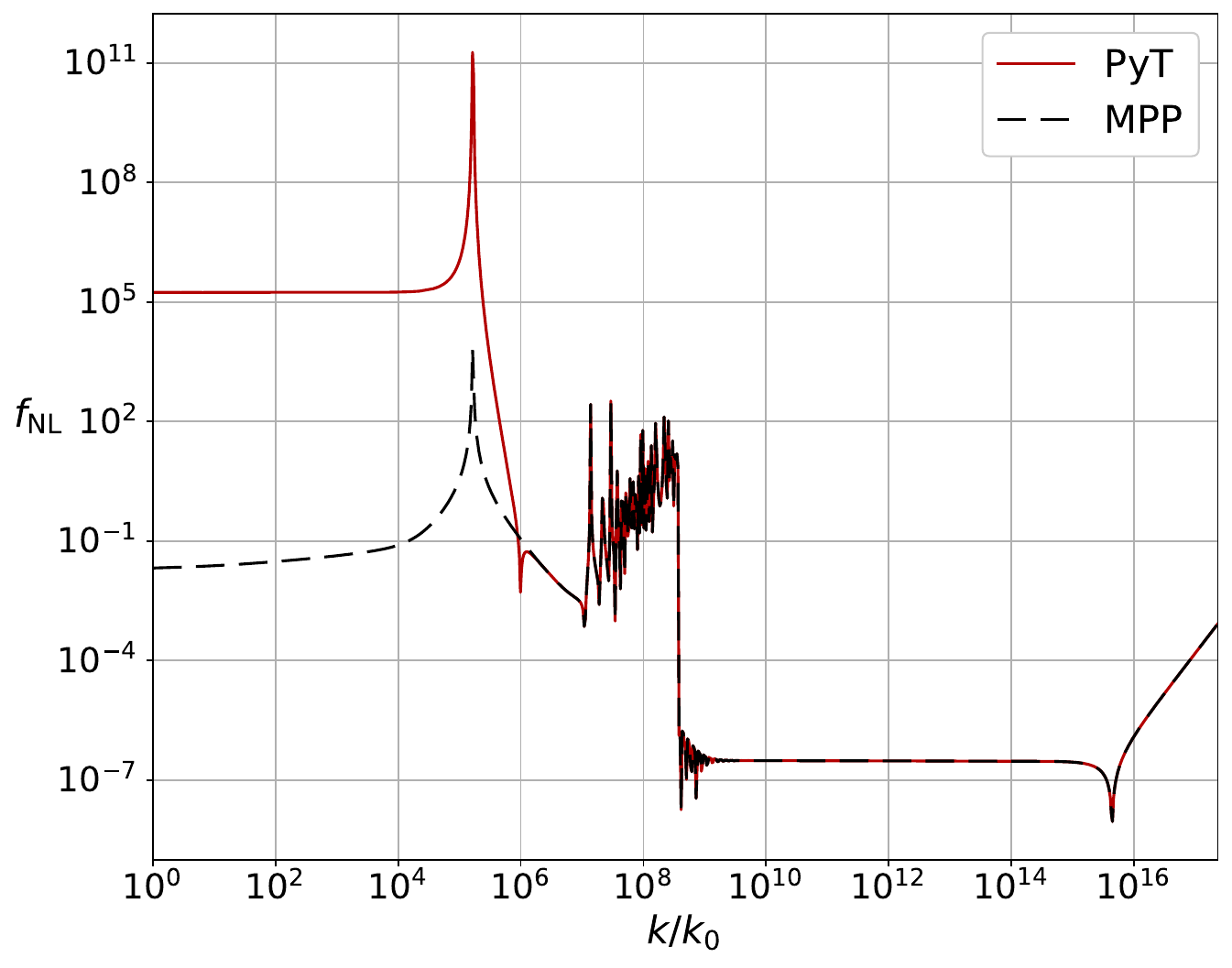}
    \end{subfigure}
    \caption{\textit{Left panel:} evolution of the phase-space 3-point correlators for a large-scale mode, $k_0$, that exits the horizon at \(N = 6\).
    Continuous (dashed) lines represent \texttt{PyTransport} (MPP) results. 
    \textit{Right panel:} reduced bispectrum in the equilateral configuration, $f_\text{NL}(k)$, computed with the two numerical approaches.}
    \label{fig: PBH Three-point and fNL}
\end{figure}

However, \texttt{PyTransport} encounters difficulties when computing non-Gaussianity for this ultra-slow-roll model. 
For modes that exit the horizon before the transition, the equilateral $f_\text{NL}(k)$ obtained with \texttt{PyTransport} is unphysically large~\cite{Maldacena:2002vr}, see the right panel of Fig.~\ref{fig: PBH Three-point and fNL}. 
This is due to the inability of \texttt{PyTransport} to precisely track the super-horizon decay of $B^{abc}$, a feature already pointed out in Sec.~\ref{sec: double quadratic}.
In the left panel of Fig.~\ref{fig: PBH Three-point and fNL}, we compare the results for $B^{abc}$ obtained from \texttt{PyTransport} against those produced within the MPP approach for a large-scale mode, $k_0$. 
The former systematically over-estimates the value of 3-point correlators at the end of inflation, and there is a seven-orders-of-magnitude discrepancy between the two codes. 
This in turns affects $f_\text{NL}(k)$, which is significantly overestimated by \texttt{PyTransport} for large-scale modes that exited the horizon before the transition. 
For shorter scales, the two codes produce the same results.

\subsection{Ultra-slow-roll model: squeezed bispectrum}
\label{sec: USR and squeezed bispectrum}

In this section we consider again a toy-model potential leading to a transient ultra-slow-roll phase (see e.g. Ref.~\cite{Iacconi:2023ggt}), 
\begin{equation}
    \frac{V\left(\phi\right)}{p_0} = p_1 + p_2\left[\ln{\left(\cosh{p_3\phi}\right)} + \left(p_3 + p_4 \right)\phi \right] \;. 
    \label{eq: SQ potential}
\end{equation}
We set $\{ p_0 = 4\times10^{-12},\, p_1 = 1,\, p_2 = 5\times10^{-7},\, p_3 = 4\times10^3,\, p_4 = 2\}$, and initial condition $\phi_\text{in} = 0.075$, with slow-roll initial velocity.
As for the model in Sec.~\ref{sec: USR and large-scale bispectrum}, we have to manually halt the evolution, and we do so after 30 e-folds. This model produces ultra-slow-roll through the potential suddenly becoming less steep -- i.e. in a different manner to the previous one since there is no ``fall''. 

\begin{figure}
    \centering
    \includegraphics[width=0.5\linewidth]{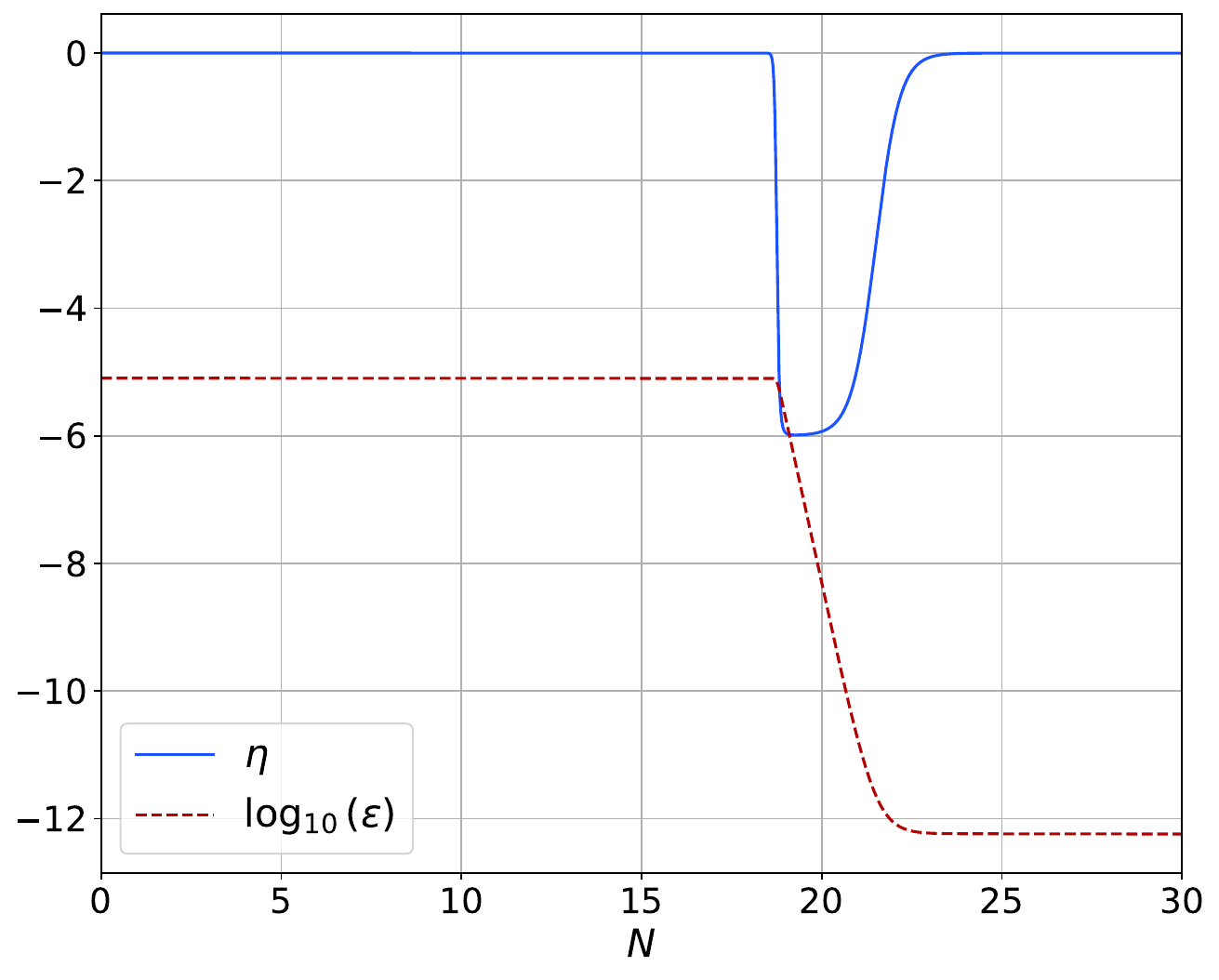}
    \caption{Time-evolution of the first two slow-roll parameters, $\epsilon$ and $\eta$, for the ultra-slow-roll model~\eqref{eq: SQ potential}.}
    \label{fig: SQ background}
\end{figure}
In Fig.~\ref{fig: SQ background} we represent the time-evolution of the first two slow-roll parameters, see Eq.~\eqref{eq: slow-roll parameter}.
Ultra-slow-roll evolution ($\eta\sim -6$) is preceded and followed by slow-roll, with smooth transitions in between. 

\begin{figure}
    \centering
    \begin{subfigure}{.45\textwidth}
        \includegraphics[width=\textwidth]{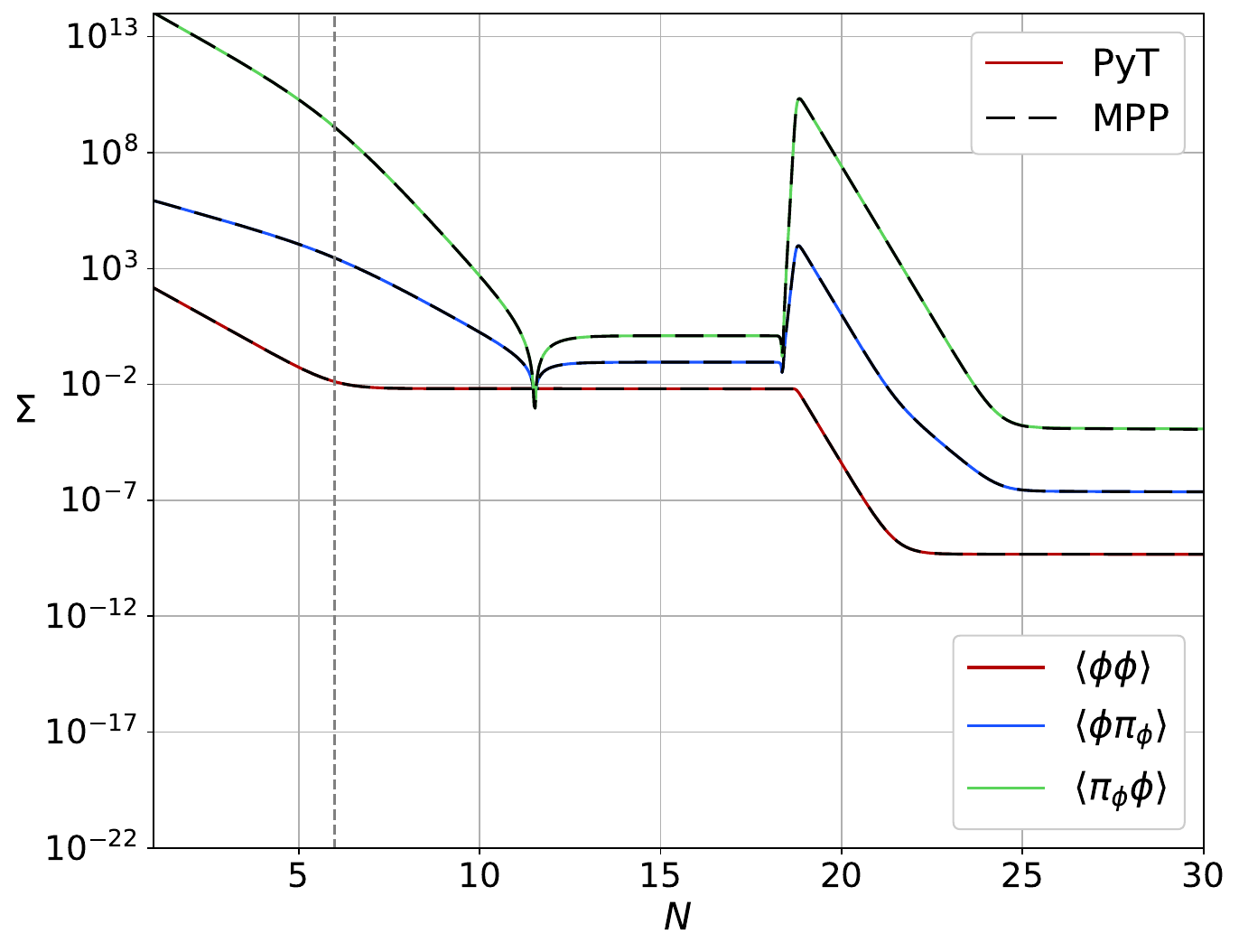}
    \end{subfigure}
    \begin{subfigure}{.45\textwidth}
        \includegraphics[width=\textwidth]{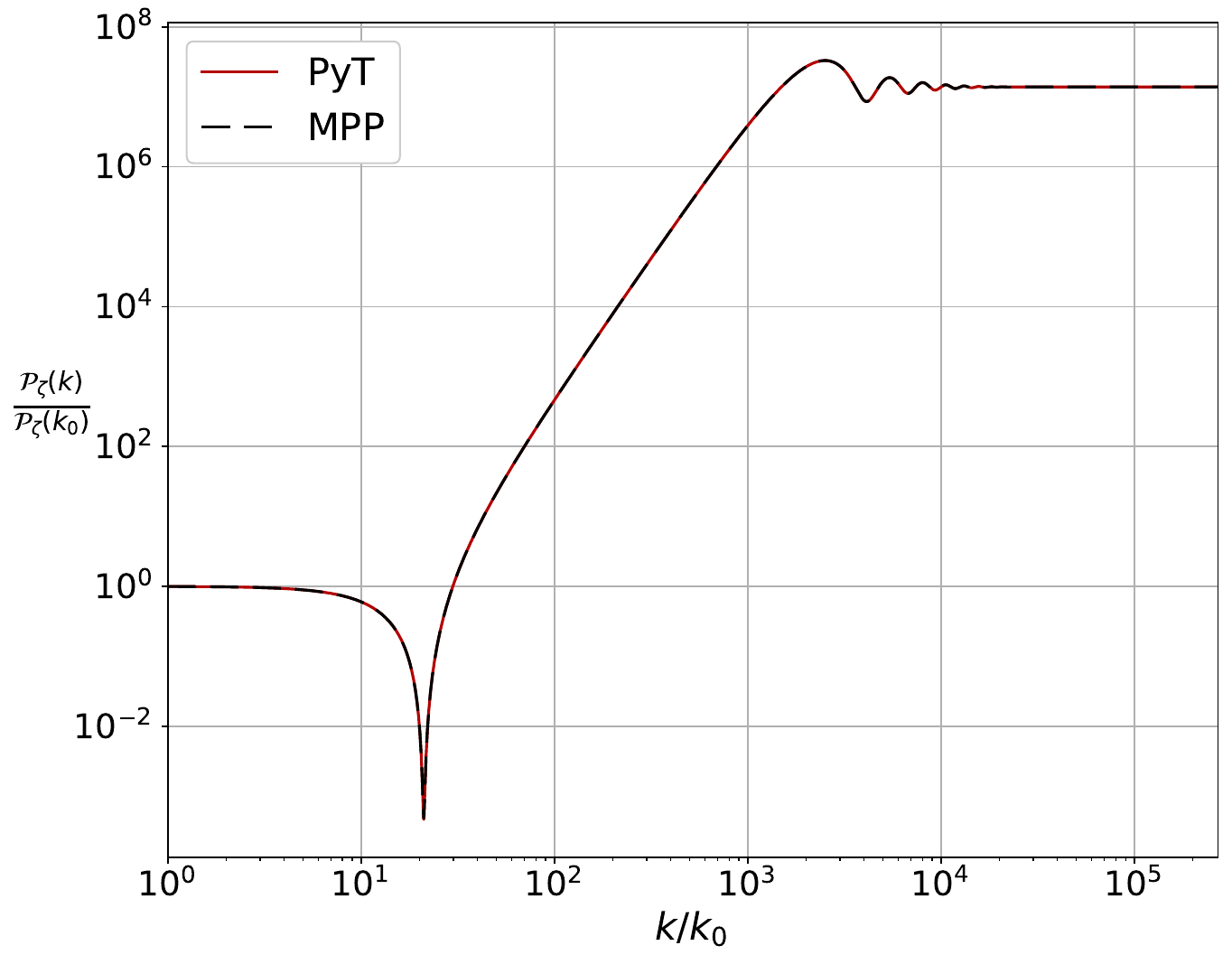}
    \end{subfigure}
    \caption{\textit{Left panel:} evolution of phase-space 2-point correlators for a large-scale mode, $k_0$, that exits the horizon at $N = 6$.
    Continuous (dashed) lines represent \texttt{PyTransport} (MPP) results. 
    \textit{Right panel:} scalar power spectrum, $\mathcal{P}_\zeta(k)$, computed with both methods.}
    \label{fig: SQ 2pt}
\end{figure}
In Fig.~\ref{fig: SQ 2pt} we display results for the time-evolution of phase-space 2-point correlators for a large-scale mode (left panel), and the primordial power spectrum $\mathcal{P}_\zeta(k)$ (right panel). 
As in Sec.~\ref{sec: USR and large-scale bispectrum} the two numerical methods show excellent agreement. 

Let us consider now the computation of the primordial bispectrum in the squeezed configuration, i.e. $f_\text{NL}(k_1,k_2,k_3)$ in the limit $k_1\ll k_2\sim k_3$. 
For ease of notation, let us identify $k_1=k_\text{long}$ and $k_2\sim k_3 = k_\text{short}$. 
\begin{figure}
    \centering
    \begin{subfigure}{.45\textwidth}
        \includegraphics[width=\textwidth]{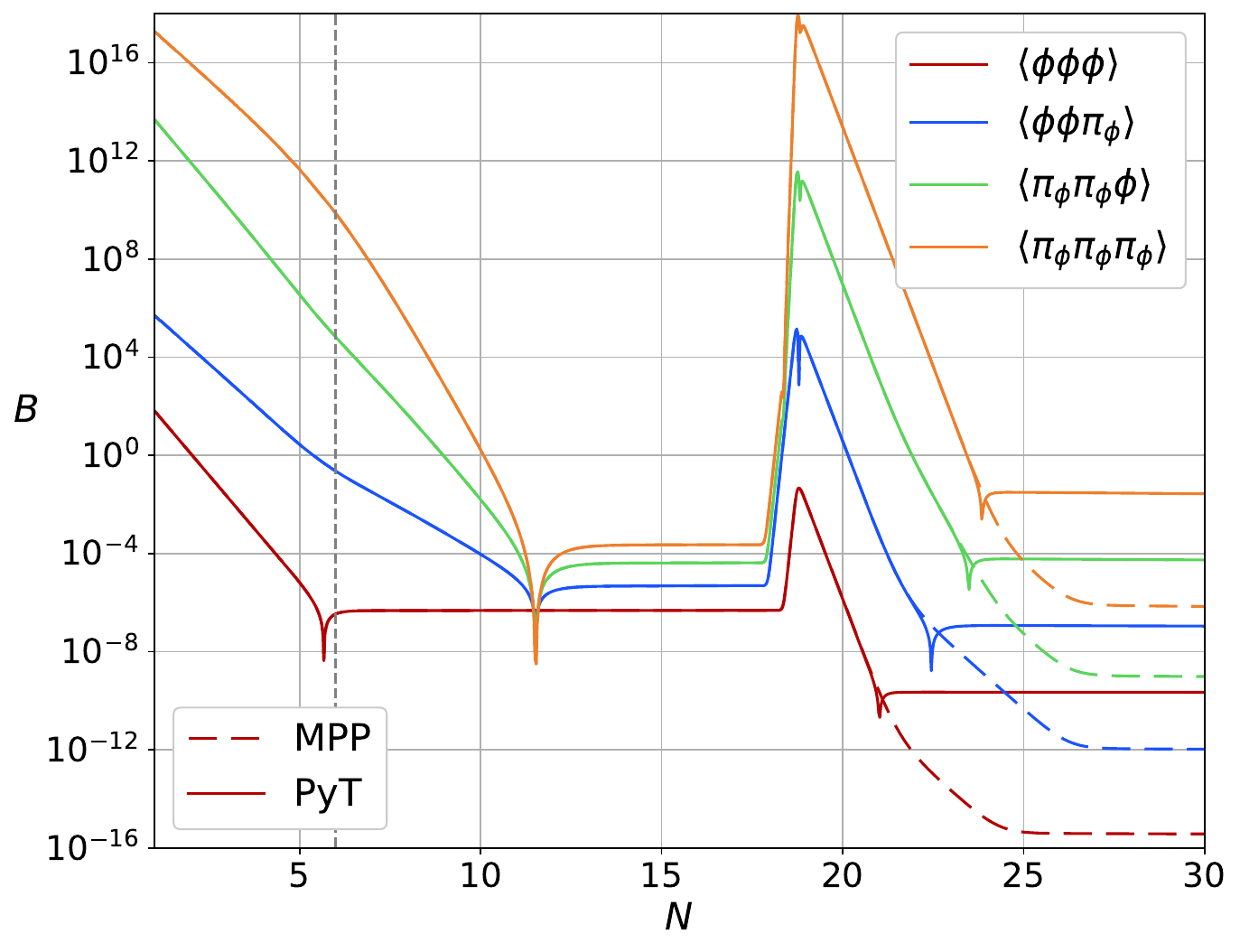}
    \end{subfigure}
    \begin{subfigure}{.45\textwidth}
        \includegraphics[width=\textwidth]{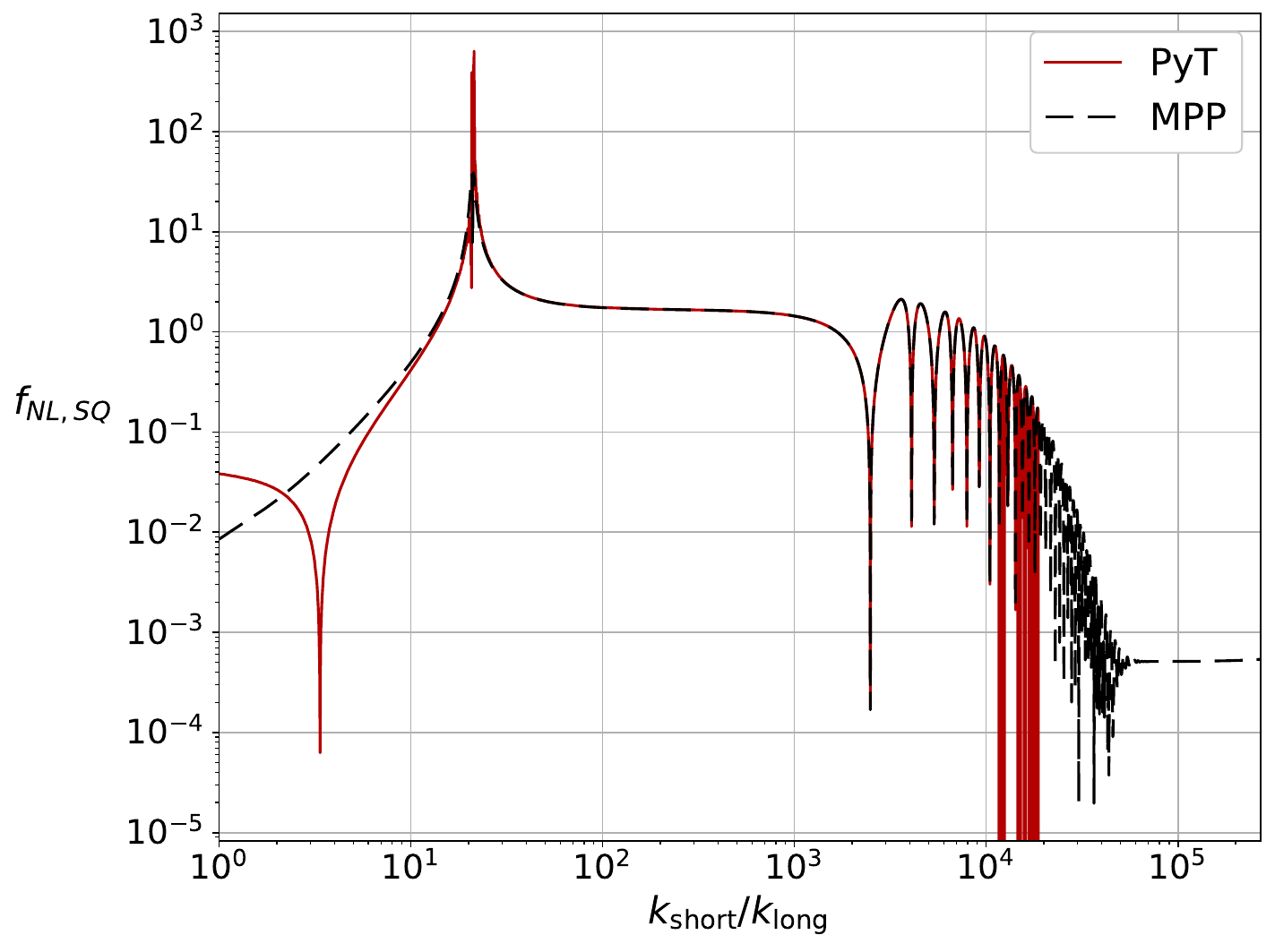}
    \end{subfigure}
    \caption{\textit{Left panel:} evolution of the phase-space 3-point correlators for a large-scale mode, $k_0$, that exits the horizon at $N = 6$. 
    Continuous (dashed) lines represent \texttt{PyTransport} (MPP) results. 
    \textit{Right panel:} squeezed non-Gaussianity, $f_\text{NL,sq}(k_\text{long},k_\text{short})$ for fixed $k_\text{long}=k_0$ as a function of $k_\text{short}/k_\text{long}$. 
    We display results computed with both methods.}
    \label{fig: SQ 3pt}
\end{figure}
In the left panel of Fig.~\ref{fig: SQ 3pt} results for the time-evolution of the phase-space 3-point correlators for a large-scale mode, $B^{abc}(k_0,k_0,k_0)$, are displayed. 
As in Sec.~\ref{sec: USR and large-scale bispectrum}, \texttt{PyTransport} consistently over-estimates the final values of these correlators, even if the discrepancy here is smaller. 
This in turns explains the results for $f_\text{NL,sq}$ represented in the right panel of Fig.~\ref{fig: SQ 3pt}. 
While the two methods agree for large squeezing values $(k_\text{short}\gg k_\text{long})$, for more modest ones, corresponding to all modes being comparably large, they disagree. 
Moreover, we note that the MPP algorithm allows computation of $f_\text{NL,sq}$ for squeezing values at least one order of magnitude larger than those attainable with \texttt{PyTransport}. 

In the squeezed regime, we can validate our numerical results by comparing them against Maldacena's consistency relation~\cite{Maldacena:2002vr} 
\begin{equation}
    f_\text{NL,sq}(k_\text{long},\, k_\text{short}) = \frac{5}{12}\left[1-n_s(k_\text{short})\right] \;. 
    \label{eq: maldacena}
\end{equation}
\begin{figure}
    \centering
    \includegraphics[width=0.9\linewidth]{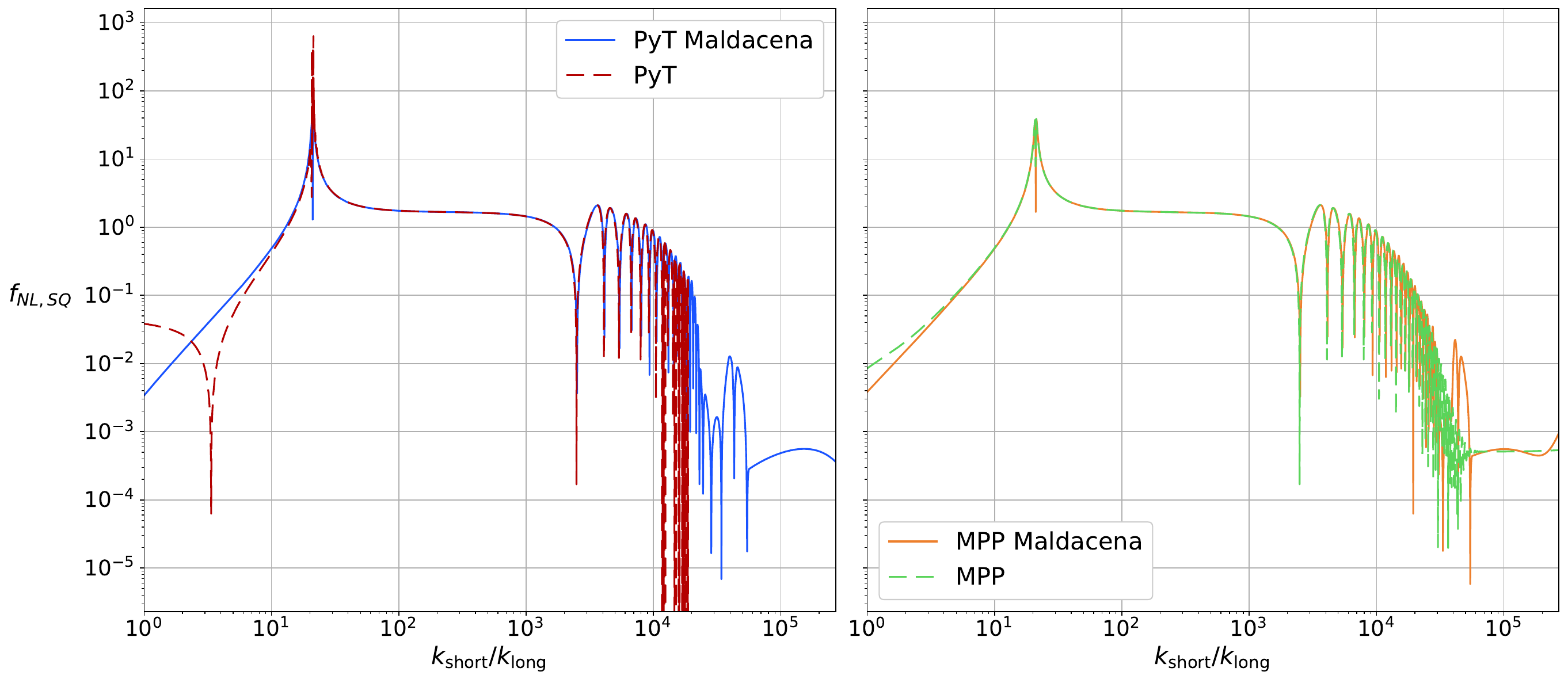}
    \caption{\textit{Left panel:} Comparison between $f_\text{NL,sq}(k_\text{long},k_\text{short})$ computed with \texttt{PyTransport} and Maldacena's consistency relation, see Eq.~\eqref{eq: maldacena}. 
    To obtain the short-scale power spectrum tilt, $n_s(k_\text{short})$, we consistently use \texttt{PyTransport} results for $\mathcal{P}_\zeta(k)$, see the red line in the right panel of Fig.~\ref{fig: SQ 2pt}.
    \textit{Right panel:} analogous comparison made for the MPP code results. }
    \label{fig: SQ Maldacena}
\end{figure}
As shown in both panels of Fig.~\ref{fig: SQ Maldacena}, the dip in $f_\text{NL,sq}$ at $k_\text{short}/k_\text{long}\sim 3$ predicted by \texttt{PyTransport} is absent in the consistency relation~\eqref{eq: maldacena}. 
The MPP algorithm produces results in agreement with the consistency relation for $3 \lesssim k_\text{short}/k_\text{long} \lesssim 2 \times 10^{4}$. For large squeezing values, finite difference methods fail to accurately compute the small-scale spectral index, and for this reason we believe that the results obtained for Eq.~\eqref{eq: maldacena} are not correct, while the MPP correctly captures the expected value of \( f_\text{NL} \).
The deviation between MPP numerical results and Eq.~\eqref{eq: maldacena} for $k_\text{short}/k_\text{long} \to 1$ is expected, as in this limit the bispectrum approaches the equilateral configuration, to which Maldacena's result does not apply. 

To summarise, our newly introduced MPP code allowed us to compute the \textit{correct} squeezed bispectrum for modest squeezing values, while 
\texttt{PyTransport}  gives an erroneous result (albeit in a regime where $f_\text{NL}$ is very small).

\section{Discussion}
\label{sec: discussion}

The transport method is a convenient framework for the numerical computation of correlators of the primordial curvature perturbation, $\zeta$. 
The (tree-level) spectrum and bispectrum of $\zeta$ are built from phase-space 2- and 3-point correlators, whose time-evolution is traditionally obtained by solving a set of appropriate differential equations, see Eqs.~\eqref{eq: traditional transport equations}.

In this work, we introduce a novel numerical implementation of the transport approach.
Key quantities are transfer ``matrices'', called \textit{multi-point propagators} (MPPs), which are momentum-dependent objects that allow the expansion of late-time phase-space variables in terms of early-time ones, see Eq.~\eqref{eq: mpp field}.
Instead of solving transport differential equations directly for phase-space correlators (as done in traditional implementations of the transport approach, such as \texttt{PyTransport}), one can set up a system of transport equations for MPPs, see Eqs.~\eqref{eq: transport eqs for MPPs}, and use the resulting MPPs to build phase-space correlators at any time of interest, see Eqs.~\eqref{eq: use MPP for sigma and B at late times}.
We implement MPPs within \texttt{PyTransport}, which allows us to take advantage of its existing architecture to treat models of inflation with multiple (scalar) fields and non-trivial field-space metric. 
The core of the MPP package is written in \texttt{C++}, with a \texttt{Python} interface. 

We test the novel MPP code in Sec.~\ref{sec: MPP at work} by comparing it with the traditional implementation of the transport approach built within \texttt{PyTransport}. 
In Secs.~\ref{sec: double quadratic}--~\ref{sec: two field non-canonical} we consider a set of inflationary models that are representative of the complexity which one might encounter in inflationary model building with scalar field(s). 
While the quantitative results of our tests are model dependent, we can infer some general features of the performance of the MPP code. 
We find that MPPs provide accurate and precise results, both for 2- and 3-point correlators. 
With MPPs we can impose the (analytical) initial conditions for the transport equations more deeply inside of the horizon than what achievable with \texttt{PyTransport}. 
This is the case for all models tested, both for the spectrum and bispectrum, making MPPs a robust tool for models where the correct handling of initial conditions requires longer sub-horizon evolution.  
When the comparison between the two codes is possible, we find that the running time of MPPs is either comparable (for computing $f_\text{NL}$) or marginally slower (for computing $\mathcal{P}_\zeta$) than that of \texttt{PyTransport}. 
For the power spectrum computation, the marginally longer running times can likely be explained in terms of tolerance settings. 
Indeed, while the MPP code is always able to reach the desired precision (even for long sub-horizon evolution), the tolerance required is smaller than what needed in \texttt{PyTransport} (due to the oscillatory nature of MPPs), and this affects the running time. 
For the bispectrum, the comparable running times might be explained by the simplicity of the equations of motions for MPPs when compared with those for the phase-space correlators, e.g. Eq.~\eqref{eq: mpp3 eom} has less terms than Eq.~\eqref{eq: transport B}. 
For this reason, MPPs might be more advantageous in computing higher-order correlators. 

For models where it is important to precisely track the super-horizon decay of correlators the MPP method proves essential.  
This is the case for the single-field models considered in Secs.~\ref{sec: USR and large-scale bispectrum} and~\ref{sec: USR and squeezed bispectrum}, which feature a transient ultra-slow-roll phase.
The modes that are most affected are large scales that crossed the horizon long before the onset of ultra-slow-roll, e.g. CMB scales. 
We find that the MPP code is able to precisely track decaying 3-point correlators, while \texttt{PyTransport} fails to track them accurately and consistently over-estimates them. 
This has important consequences for the curvature bispectrum on large scales. 
In Sec.~\ref{sec: USR and large-scale bispectrum} we show that while \texttt{PyTransport} wrongly predicts $f_\text{NL}\sim \mathcal{O}(10^5)$, the expected slow-roll-suppressed value~\cite{Maldacena:2002vr} is recovered with the MPP code, $f_\text{NL}\sim \mathcal{O}(10^{-2})$. 
In Sec.~\ref{sec: USR and squeezed bispectrum} we find similar results for squeezed bispectra when the squeezing is modest. 
In addition, we find that MPPs allow to extend the computation of the squeezed bispectrum for squeezing values at least one decade beyond those attainable with \texttt{PyTransport}. 

We conclude by mentioning a future interesting application of MPPs. 
Correlators beyond tree-level, e.g. at 1-loop, can be built from MPPs~\cite{Iacconi_in_preparation}. The expressions for the loops involved are presented in Ref.~\cite{Iacconi_in_preparation}, where they are used for analytic purposes.  In combination with our present numerical work, however, this could also allow the efficient numerical calculation of primordial correlators at 1-loop. 
We will return to this application of the MPP formalism in future work.  

\section*{Acknowledgments}
The authors would like to thank David Seery for many interesting and useful discussions. 
AC is supported by a studentship awarded by the Perren Bequest. 
LI was supported by a Royal Society funded post-doctoral position, and acknowledges current financial support from the STFC under grant ST/X000931/1.
DJM was supported by a Royal Society University Research Fellowship for the majority of this work, and acknowledges current financial support from the STFC under grant ST/X000931/1.

\bibliography{refs} 

\providecommand{\href}[2]{#2}\begingroup\raggedright\begin{thebibliography}{10}

\bibitem{Starobinsky:1980te}
A.~A. Starobinsky, \emph{{A New Type of Isotropic Cosmological Models Without Singularity}}, \href{https://doi.org/10.1016/0370-2693(80)90670-X}{\emph{Phys. Lett. B} {\bfseries 91} (1980) 99}.

\bibitem{PhysRevD.23.347}
A.~H. Guth, \emph{Inflationary universe: A possible solution to the horizon and flatness problems}, \href{https://doi.org/10.1103/PhysRevD.23.347}{\emph{Phys. Rev. D} {\bfseries 23} (1981) 347}.

\bibitem{PhysRevLett.48.1220}
A.~Albrecht and P.~J. Steinhardt, \emph{Cosmology for grand unified theories with radiatively induced symmetry breaking}, \href{https://doi.org/10.1103/PhysRevLett.48.1220}{\emph{Phys. Rev. Lett.} {\bfseries 48} (1982) 1220}.

\bibitem{Hawking:1981fz}
S.~W. Hawking and I.~G. Moss, \emph{{Supercooled Phase Transitions in the Very Early Universe}}, \href{https://doi.org/10.1016/0370-2693(82)90946-7}{\emph{Phys. Lett. B} {\bfseries 110} (1982) 35}.

\bibitem{Linde:1981mu}
A.~D. Linde, \emph{{A New Inflationary Universe Scenario: A Possible Solution of the Horizon, Flatness, Homogeneity, Isotropy and Primordial Monopole Problems}}, \href{https://doi.org/10.1016/0370-2693(82)91219-9}{\emph{Phys. Lett. B} {\bfseries 108} (1982) 389}.

\bibitem{Linde:1983gd}
A.~D. Linde, \emph{{Chaotic Inflation}}, \href{https://doi.org/10.1016/0370-2693(83)90837-7}{\emph{Phys. Lett. B} {\bfseries 129} (1983) 177}.

\bibitem{Liddle_Lyth_2000}
A.~R. Liddle and D.~H. Lyth, \emph{Cosmological Inflation and Large-Scale Structure}. Cambridge University Press, 2000.

\bibitem{Planck:2018jri}
{\scshape Planck} collaboration, \emph{{Planck 2018 results. X. Constraints on inflation}}, \href{https://doi.org/10.1051/0004-6361/201833887}{\emph{Astron. Astrophys.} {\bfseries 641} (2020) A10} [\href{https://arxiv.org/abs/1807.06211}{{\ttfamily 1807.06211}}].

\bibitem{Gangui:1993tt}
A.~Gangui, F.~Lucchin, S.~Matarrese and S.~Mollerach, \emph{{The Three point correlation function of the cosmic microwave background in inflationary models}}, \href{https://doi.org/10.1086/174421}{\emph{Astrophys. J.} {\bfseries 430} (1994) 447} [\href{https://arxiv.org/abs/astro-ph/9312033}{{\ttfamily astro-ph/9312033}}].

\bibitem{Maldacena:2002vr}
J.~M. Maldacena, \emph{{Non-Gaussian features of primordial fluctuations in single field inflationary models}}, \href{https://doi.org/10.1088/1126-6708/2003/05/013}{\emph{JHEP} {\bfseries 05} (2003) 013} [\href{https://arxiv.org/abs/astro-ph/0210603}{{\ttfamily astro-ph/0210603}}].

\bibitem{Celoria:2018euj}
M.~Celoria and S.~Matarrese, \emph{{Primordial Non-Gaussianity}}, \href{https://doi.org/10.3254/ENFI200009}{\emph{Proc. Int. Sch. Phys. Fermi} {\bfseries 200} (2020) 179} [\href{https://arxiv.org/abs/1812.08197}{{\ttfamily 1812.08197}}].

\bibitem{Planck:2019kim}
{\scshape Planck} collaboration, \emph{{Planck 2018 results. IX. Constraints on primordial non-Gaussianity}}, \href{https://doi.org/10.1051/0004-6361/201935891}{\emph{Astron. Astrophys.} {\bfseries 641} (2020) A9} [\href{https://arxiv.org/abs/1905.05697}{{\ttfamily 1905.05697}}].

\bibitem{Achucarro:2022qrl}
A.~Ach\'ucarro et~al., \emph{{Inflation: Theory and Observations}},  \href{https://arxiv.org/abs/2203.08128}{{\ttfamily 2203.08128}}.

\bibitem{Gow:2020bzo}
A.~D. Gow, C.~T. Byrnes, P.~S. Cole and S.~Young, \emph{{The power spectrum on small scales: Robust constraints and comparing PBH methodologies}}, \href{https://doi.org/10.1088/1475-7516/2021/02/002}{\emph{JCAP} {\bfseries 02} (2021) 002} [\href{https://arxiv.org/abs/2008.03289}{{\ttfamily 2008.03289}}].

\bibitem{LISACosmologyWorkingGroup:2025vdz}
{\scshape LISA Cosmology Working Group} collaboration, \emph{{Reconstructing Primordial Curvature Perturbations via Scalar-Induced Gravitational Waves with LISA}},  \href{https://arxiv.org/abs/2501.11320}{{\ttfamily 2501.11320}}.

\bibitem{Carr:1974nx}
B.~J. Carr and S.~W. Hawking, \emph{{Black holes in the early Universe}}, \href{https://doi.org/10.1093/mnras/168.2.399}{\emph{Mon. Not. Roy. Astron. Soc.} {\bfseries 168} (1974) 399}.

\bibitem{Green:2024bam}
A.~M. Green, \emph{{Primordial black holes as a dark matter candidate - a brief overview}}, \href{https://doi.org/10.1016/j.nuclphysb.2024.116494}{\emph{Nucl. Phys. B} {\bfseries 1003} (2024) 116494} [\href{https://arxiv.org/abs/2402.15211}{{\ttfamily 2402.15211}}].

\bibitem{Ferrante:2022mui}
G.~Ferrante, G.~Franciolini, A.~Iovino, Junior. and A.~Urbano, \emph{{Primordial non-Gaussianity up to all orders: Theoretical aspects and implications for primordial black hole models}}, \href{https://doi.org/10.1103/PhysRevD.107.043520}{\emph{Phys. Rev. D} {\bfseries 107} (2023) 043520} [\href{https://arxiv.org/abs/2211.01728}{{\ttfamily 2211.01728}}].

\bibitem{Gow:2022jfb}
A.~D. Gow, H.~Assadullahi, J.~H.~P. Jackson, K.~Koyama, V.~Vennin and D.~Wands, \emph{{Non-perturbative non-Gaussianity and primordial black holes}},  \href{https://arxiv.org/abs/2211.08348}{{\ttfamily 2211.08348}}.

\bibitem{Guzzetti:2016mkm}
M.~C. Guzzetti, N.~Bartolo, M.~Liguori and S.~Matarrese, \emph{{Gravitational waves from inflation}}, \href{https://doi.org/10.1393/ncr/i2016-10127-1}{\emph{Riv. Nuovo Cim.} {\bfseries 39} (2016) 399} [\href{https://arxiv.org/abs/1605.01615}{{\ttfamily 1605.01615}}].

\bibitem{Baumann:2014nda}
D.~Baumann and L.~McAllister, \emph{{Inflation and String Theory}}, Cambridge Monographs on Mathematical Physics. Cambridge University Press, 5, 2015, \href{https://doi.org/10.1017/CBO9781316105733}{10.1017/CBO9781316105733}, [\href{https://arxiv.org/abs/1404.2601}{{\ttfamily 1404.2601}}].

\bibitem{Birrell:1982ix}
N.~D. Birrell and P.~C.~W. Davies, \emph{{Quantum Fields in Curved Space}}, Cambridge Monographs on Mathematical Physics. Cambridge Univ. Press, Cambridge, UK, 2, 1984, \href{https://doi.org/10.1017/CBO9780511622632}{10.1017/CBO9780511622632}.

\bibitem{Weinberg:2005vy}
S.~Weinberg, \emph{{Quantum contributions to cosmological correlations}}, \href{https://doi.org/10.1103/PhysRevD.72.043514}{\emph{Phys. Rev. D} {\bfseries 72} (2005) 043514} [\href{https://arxiv.org/abs/hep-th/0506236}{{\ttfamily hep-th/0506236}}].

\bibitem{Weinberg:2006ac}
S.~Weinberg, \emph{{Quantum contributions to cosmological correlations. II. Can these corrections become large?}}, \href{https://doi.org/10.1103/PhysRevD.74.023508}{\emph{Phys. Rev. D} {\bfseries 74} (2006) 023508} [\href{https://arxiv.org/abs/hep-th/0605244}{{\ttfamily hep-th/0605244}}].

\bibitem{Seery:2005wm}
D.~Seery and J.~E. Lidsey, \emph{{Primordial non-Gaussianities in single field inflation}}, \href{https://doi.org/10.1088/1475-7516/2005/06/003}{\emph{JCAP} {\bfseries 06} (2005) 003} [\href{https://arxiv.org/abs/astro-ph/0503692}{{\ttfamily astro-ph/0503692}}].

\bibitem{Seery:2005gb}
D.~Seery and J.~E. Lidsey, \emph{{Primordial non-Gaussianities from multiple-field inflation}}, \href{https://doi.org/10.1088/1475-7516/2005/09/011}{\emph{JCAP} {\bfseries 09} (2005) 011} [\href{https://arxiv.org/abs/astro-ph/0506056}{{\ttfamily astro-ph/0506056}}].

\bibitem{Hazra:2012yn}
D.~K. Hazra, L.~Sriramkumar and J.~Martin, \emph{{BINGO: A code for the efficient computation of the scalar bi-spectrum}}, \href{https://doi.org/10.1088/1475-7516/2013/05/026}{\emph{JCAP} {\bfseries 05} (2013) 026} [\href{https://arxiv.org/abs/1201.0926}{{\ttfamily 1201.0926}}].

\bibitem{Mulryne:2009kh}
D.~J. Mulryne, D.~Seery and D.~Wesley, \emph{{Moment transport equations for non-Gaussianity}}, \href{https://doi.org/10.1088/1475-7516/2010/01/024}{\emph{JCAP} {\bfseries 01} (2010) 024} [\href{https://arxiv.org/abs/0909.2256}{{\ttfamily 0909.2256}}].

\bibitem{Mulryne:2010rp}
D.~J. Mulryne, D.~Seery and D.~Wesley, \emph{{Moment transport equations for the primordial curvature perturbation}}, \href{https://doi.org/10.1088/1475-7516/2011/04/030}{\emph{JCAP} {\bfseries 04} (2011) 030} [\href{https://arxiv.org/abs/1008.3159}{{\ttfamily 1008.3159}}].

\bibitem{Seery:2012vj}
D.~Seery, D.~J. Mulryne, J.~Frazer and R.~H. Ribeiro, \emph{{Inflationary perturbation theory is geometrical optics in phase space}}, \href{https://doi.org/10.1088/1475-7516/2012/09/010}{\emph{JCAP} {\bfseries 09} (2012) 010} [\href{https://arxiv.org/abs/1203.2635}{{\ttfamily 1203.2635}}].

\bibitem{Anderson:2012em}
G.~J. Anderson, D.~J. Mulryne and D.~Seery, \emph{{Transport equations for the inflationary trispectrum}}, \href{https://doi.org/10.1088/1475-7516/2012/10/019}{\emph{JCAP} {\bfseries 10} (2012) 019} [\href{https://arxiv.org/abs/1205.0024}{{\ttfamily 1205.0024}}].

\bibitem{Mulryne:2013uka}
D.~J. Mulryne, \emph{{Transporting non-Gaussianity from sub to super-horizon scales}}, \href{https://doi.org/10.1088/1475-7516/2013/09/010}{\emph{JCAP} {\bfseries 09} (2013) 010} [\href{https://arxiv.org/abs/1302.3842}{{\ttfamily 1302.3842}}].

\bibitem{Dias:2016rjq}
M.~Dias, J.~Frazer, D.~J. Mulryne and D.~Seery, \emph{{Numerical evaluation of the bispectrum in multiple field inflation\textemdash{}the transport approach with code}}, \href{https://doi.org/10.1088/1475-7516/2016/12/033}{\emph{JCAP} {\bfseries 12} (2016) 033} [\href{https://arxiv.org/abs/1609.00379}{{\ttfamily 1609.00379}}].

\bibitem{Dias:2015rca}
M.~Dias, J.~Frazer and D.~Seery, \emph{{Computing observables in curved multifield models of inflation\textemdash{}A guide (with code) to the transport method}}, \href{https://doi.org/10.1088/1475-7516/2015/12/030}{\emph{JCAP} {\bfseries 12} (2015) 030} [\href{https://arxiv.org/abs/1502.03125}{{\ttfamily 1502.03125}}].

\bibitem{Seery:2016lko}
D.~Seery, \emph{{CppTransport: a platform to automate calculation of inflationary correlation functions}},  \href{https://arxiv.org/abs/1609.00380}{{\ttfamily 1609.00380}}.

\bibitem{Mulryne:2016mzv}
D.~J. Mulryne and J.~W. Ronayne, \emph{{PyTransport: A Python package for the calculation of inflationary correlation functions}}, \href{https://doi.org/10.21105/joss.00494}{\emph{J. Open Source Softw.} {\bfseries 3} (2018) 494} [\href{https://arxiv.org/abs/1609.00381}{{\ttfamily 1609.00381}}].

\bibitem{Ronayne:2017qzn}
J.~W. Ronayne and D.~J. Mulryne, \emph{{Numerically evaluating the bispectrum in curved field-space\textemdash{} with PyTransport 2.0}}, \href{https://doi.org/10.1088/1475-7516/2018/01/023}{\emph{JCAP} {\bfseries 01} (2018) 023} [\href{https://arxiv.org/abs/1708.07130}{{\ttfamily 1708.07130}}].

\bibitem{Werth:2023pfl}
D.~Werth, L.~Pinol and S.~Renaux-Petel, \emph{{Cosmological Flow of Primordial Correlators}}, \href{https://doi.org/10.1103/PhysRevLett.133.141002}{\emph{Phys. Rev. Lett.} {\bfseries 133} (2024) 141002} [\href{https://arxiv.org/abs/2302.00655}{{\ttfamily 2302.00655}}].

\bibitem{Pinol:2023oux}
L.~Pinol, S.~Renaux-Petel and D.~Werth, \emph{{The Cosmological Flow: A Systematic Approach to Primordial Correlators}},  \href{https://arxiv.org/abs/2312.06559}{{\ttfamily 2312.06559}}.

\bibitem{Werth:2024aui}
D.~Werth, L.~Pinol and S.~Renaux-Petel, \emph{{${\definecolor{pyblue}{RGB}{31, 119, 180}\color{pyblue}{\mathcal{C}}}\texttt{osmo}{\definecolor{pyred}{RGB}{214, 39, 40}\color{pyred}{\mathcal{F}}}\texttt{low}$: Python package for cosmological correlators}}, \href{https://doi.org/10.1088/1361-6382/ad6740}{\emph{Class. Quant. Grav.} {\bfseries 41} (2024) 175015} [\href{https://arxiv.org/abs/2402.03693}{{\ttfamily 2402.03693}}].

\bibitem{Bernardeau:2008fa}
F.~Bernardeau, M.~Crocce and R.~Scoccimarro, \emph{{Multi-Point Propagators in Cosmological Gravitational Instability}}, \href{https://doi.org/10.1103/PhysRevD.78.103521}{\emph{Phys. Rev. D} {\bfseries 78} (2008) 103521} [\href{https://arxiv.org/abs/0806.2334}{{\ttfamily 0806.2334}}].

\bibitem{Kinney:2005vj}
W.~H. Kinney, \emph{{Horizon crossing and inflation with large eta}}, \href{https://doi.org/10.1103/PhysRevD.72.023515}{\emph{Phys. Rev. D} {\bfseries 72} (2005) 023515} [\href{https://arxiv.org/abs/gr-qc/0503017}{{\ttfamily gr-qc/0503017}}].

\bibitem{Dimopoulos:2017ged}
K.~Dimopoulos, \emph{{Ultra slow-roll inflation demystified}}, \href{https://doi.org/10.1016/j.physletb.2017.10.066}{\emph{Phys. Lett. B} {\bfseries 775} (2017) 262} [\href{https://arxiv.org/abs/1707.05644}{{\ttfamily 1707.05644}}].

\bibitem{Pattison:2018bct}
C.~Pattison, V.~Vennin, H.~Assadullahi and D.~Wands, \emph{{The attractive behaviour of ultra-slow-roll inflation}}, \href{https://doi.org/10.1088/1475-7516/2018/08/048}{\emph{JCAP} {\bfseries 08} (2018) 048} [\href{https://arxiv.org/abs/1806.09553}{{\ttfamily 1806.09553}}].

\bibitem{Costantini_in_prep}
A.~Costantini, L.~Iacconi and D.~J. Mulryne, \emph{{The MPP package in PyTransport}}, {\emph{to appear} }.

\bibitem{Gong:2011uw}
J.-O. Gong and T.~Tanaka, \emph{{A covariant approach to general field space metric in multi-field inflation}}, \href{https://doi.org/10.1088/1475-7516/2012/02/E01}{\emph{JCAP} {\bfseries 03} (2011) 015} [\href{https://arxiv.org/abs/1101.4809}{{\ttfamily 1101.4809}}].

\bibitem{Arnowitt:1962hi}
R.~L. Arnowitt, S.~Deser and C.~W. Misner, \emph{{The Dynamics of general relativity}}, \href{https://doi.org/10.1007/s10714-008-0661-1}{\emph{Gen. Rel. Grav.} {\bfseries 40} (2008) 1997} [\href{https://arxiv.org/abs/gr-qc/0405109}{{\ttfamily gr-qc/0405109}}].

\bibitem{Misner:1973prb}
C.~W. Misner, K.~S. Thorne and J.~A. Wheeler, \emph{{Gravitation}}. W. H. Freeman, San Francisco, 1973.

\bibitem{Dias:2014msa}
M.~Dias, J.~Elliston, J.~Frazer, D.~Mulryne and D.~Seery, \emph{{The curvature perturbation at second order}}, \href{https://doi.org/10.1088/1475-7516/2015/02/040}{\emph{JCAP} {\bfseries 02} (2015) 040} [\href{https://arxiv.org/abs/1410.3491}{{\ttfamily 1410.3491}}].

\bibitem{LISACosmologyWorkingGroup:2024hsc}
{\scshape LISA Cosmology Working Group} collaboration, \emph{{Gravitational waves from inflation in LISA: reconstruction pipeline and physics interpretation}}, \href{https://doi.org/10.1088/1475-7516/2024/11/032}{\emph{JCAP} {\bfseries 11} (2024) 032} [\href{https://arxiv.org/abs/2407.04356}{{\ttfamily 2407.04356}}].

\bibitem{Ozsoy:2023ryl}
O.~\"Ozsoy and G.~Tasinato, \emph{{Inflation and Primordial Black Holes}}, \href{https://doi.org/10.3390/universe9050203}{\emph{Universe} {\bfseries 9} (2023) 203} [\href{https://arxiv.org/abs/2301.03600}{{\ttfamily 2301.03600}}].

\bibitem{Silk:1986vc}
J.~Silk and M.~S. Turner, \emph{{Double Inflation}}, \href{https://doi.org/10.1103/PhysRevD.35.419}{\emph{Phys. Rev. D} {\bfseries 35} (1987) 419}.

\bibitem{Polarski:1992dq}
D.~Polarski and A.~A. Starobinsky, \emph{{Spectra of perturbations produced by double inflation with an intermediate matter dominated stage}}, \href{https://doi.org/10.1016/0550-3213(92)90062-G}{\emph{Nucl. Phys. B} {\bfseries 385} (1992) 623}.

\bibitem{Polarski:1994rz}
D.~Polarski and A.~A. Starobinsky, \emph{{Isocurvature perturbations in multiple inflationary models}}, \href{https://doi.org/10.1103/PhysRevD.50.6123}{\emph{Phys. Rev. D} {\bfseries 50} (1994) 6123} [\href{https://arxiv.org/abs/astro-ph/9404061}{{\ttfamily astro-ph/9404061}}].

\bibitem{Langlois:1999dw}
D.~Langlois, \emph{{Correlated adiabatic and isocurvature perturbations from double inflation}}, \href{https://doi.org/10.1103/PhysRevD.59.123512}{\emph{Phys. Rev. D} {\bfseries 59} (1999) 123512} [\href{https://arxiv.org/abs/astro-ph/9906080}{{\ttfamily astro-ph/9906080}}].

\bibitem{Vernizzi:2006ve}
F.~Vernizzi and D.~Wands, \emph{{Non-gaussianities in two-field inflation}}, \href{https://doi.org/10.1088/1475-7516/2006/05/019}{\emph{JCAP} {\bfseries 05} (2006) 019} [\href{https://arxiv.org/abs/astro-ph/0603799}{{\ttfamily astro-ph/0603799}}].

\bibitem{Martin:2014nya}
J.~Martin, C.~Ringeval and V.~Vennin, \emph{{Observing Inflationary Reheating}}, \href{https://doi.org/10.1103/PhysRevLett.114.081303}{\emph{Phys. Rev. Lett.} {\bfseries 114} (2015) 081303} [\href{https://arxiv.org/abs/1410.7958}{{\ttfamily 1410.7958}}].

\bibitem{Dimopoulos:2005ac}
S.~Dimopoulos, S.~Kachru, J.~McGreevy and J.~G. Wacker, \emph{{N-flation}}, \href{https://doi.org/10.1088/1475-7516/2008/08/003}{\emph{JCAP} {\bfseries 08} (2008) 003} [\href{https://arxiv.org/abs/hep-th/0507205}{{\ttfamily hep-th/0507205}}].

\bibitem{Elliston:2012wm}
J.~Elliston, L.~Alabidi, I.~Huston, D.~Mulryne and R.~Tavakol, \emph{{Large trispectrum in two-field slow-roll inflation}}, \href{https://doi.org/10.1088/1475-7516/2012/09/001}{\emph{JCAP} {\bfseries 09} (2012) 001} [\href{https://arxiv.org/abs/1203.6844}{{\ttfamily 1203.6844}}].

\bibitem{Kim:2010ud}
S.~A. Kim, A.~R. Liddle and D.~Seery, \emph{{Non-gaussianity in axion Nflation models}}, \href{https://doi.org/10.1103/PhysRevLett.105.181302}{\emph{Phys. Rev. Lett.} {\bfseries 105} (2010) 181302} [\href{https://arxiv.org/abs/1005.4410}{{\ttfamily 1005.4410}}].

\bibitem{Kim:2011jea}
S.~A. Kim, A.~R. Liddle and D.~Seery, \emph{{Non-gaussianity in axion N-flation models: detailed predictions and mass spectra}}, \href{https://doi.org/10.1103/PhysRevD.85.023532}{\emph{Phys. Rev. D} {\bfseries 85} (2012) 023532} [\href{https://arxiv.org/abs/1108.2944}{{\ttfamily 1108.2944}}].

\bibitem{Elliston:2011dr}
J.~Elliston, D.~J. Mulryne, D.~Seery and R.~Tavakol, \emph{{Evolution of fNL to the adiabatic limit}}, \href{https://doi.org/10.1088/1475-7516/2011/11/005}{\emph{JCAP} {\bfseries 11} (2011) 005} [\href{https://arxiv.org/abs/1106.2153}{{\ttfamily 1106.2153}}].

\bibitem{Adams:2001vc}
J.~A. Adams, B.~Cresswell and R.~Easther, \emph{{Inflationary perturbations from a potential with a step}}, \href{https://doi.org/10.1103/PhysRevD.64.123514}{\emph{Phys. Rev. D} {\bfseries 64} (2001) 123514} [\href{https://arxiv.org/abs/astro-ph/0102236}{{\ttfamily astro-ph/0102236}}].

\bibitem{Chen:2006xjb}
X.~Chen, R.~Easther and E.~A. Lim, \emph{{Large Non-Gaussianities in Single Field Inflation}}, \href{https://doi.org/10.1088/1475-7516/2007/06/023}{\emph{JCAP} {\bfseries 06} (2007) 023} [\href{https://arxiv.org/abs/astro-ph/0611645}{{\ttfamily astro-ph/0611645}}].

\bibitem{Chen:2008wn}
X.~Chen, R.~Easther and E.~A. Lim, \emph{{Generation and Characterization of Large Non-Gaussianities in Single Field Inflation}}, \href{https://doi.org/10.1088/1475-7516/2008/04/010}{\emph{JCAP} {\bfseries 04} (2008) 010} [\href{https://arxiv.org/abs/0801.3295}{{\ttfamily 0801.3295}}].

\bibitem{Braglia:2020eai}
M.~Braglia, D.~K. Hazra, F.~Finelli, G.~F. Smoot, L.~Sriramkumar and A.~A. Starobinsky, \emph{{Generating PBHs and small-scale GWs in two-field models of inflation}}, \href{https://doi.org/10.1088/1475-7516/2020/08/001}{\emph{JCAP} {\bfseries 08} (2020) 001} [\href{https://arxiv.org/abs/2005.02895}{{\ttfamily 2005.02895}}].

\bibitem{Iacconi:2023slv}
L.~Iacconi and D.~J. Mulryne, \emph{{Multi-field inflation with large scalar fluctuations: non-Gaussianity and perturbativity}}, \href{https://doi.org/10.1088/1475-7516/2023/09/033}{\emph{JCAP} {\bfseries 09} (2023) 033} [\href{https://arxiv.org/abs/2304.14260}{{\ttfamily 2304.14260}}].

\bibitem{Atal:2018neu}
V.~Atal and C.~Germani, \emph{{The role of non-gaussianities in Primordial Black Hole formation}}, \href{https://doi.org/10.1016/j.dark.2019.100275}{\emph{Phys. Dark Univ.} {\bfseries 24} (2019) 100275} [\href{https://arxiv.org/abs/1811.07857}{{\ttfamily 1811.07857}}].

\bibitem{Inomata:2021uqj}
K.~Inomata, E.~McDonough and W.~Hu, \emph{{Primordial black holes arise when the inflaton falls}}, \href{https://doi.org/10.1103/PhysRevD.104.123553}{\emph{Phys. Rev. D} {\bfseries 104} (2021) 123553} [\href{https://arxiv.org/abs/2104.03972}{{\ttfamily 2104.03972}}].

\bibitem{Carrilho:2019oqg}
P.~Carrilho, K.~A. Malik and D.~J. Mulryne, \emph{{Dissecting the growth of the power spectrum for primordial black holes}}, \href{https://doi.org/10.1103/PhysRevD.100.103529}{\emph{Phys. Rev. D} {\bfseries 100} (2019) 103529} [\href{https://arxiv.org/abs/1907.05237}{{\ttfamily 1907.05237}}].

\bibitem{Hertzberg:2017dkh}
M.~P. Hertzberg and M.~Yamada, \emph{{Primordial Black Holes from Polynomial Potentials in Single Field Inflation}}, \href{https://doi.org/10.1103/PhysRevD.97.083509}{\emph{Phys. Rev. D} {\bfseries 97} (2018) 083509} [\href{https://arxiv.org/abs/1712.09750}{{\ttfamily 1712.09750}}].

\bibitem{Iacconi:2023ggt}
L.~Iacconi, D.~Mulryne and D.~Seery, \emph{{Loop corrections in the separate universe picture}}, \href{https://doi.org/10.1088/1475-7516/2024/06/062}{\emph{JCAP} {\bfseries 06} (2024) 062} [\href{https://arxiv.org/abs/2312.12424}{{\ttfamily 2312.12424}}].

\bibitem{Iacconi_in_preparation}
L.~Iacconi, D.~J. Mulryne and D.~Seery, \emph{{Mapping inflationary loop corrections to boundary terms}}, {\emph{to appear} }.

\end{thebibliography}\endgroup

\bibliographystyle{JHEP}

\end{document}